\documentclass[10pt,conference]{IEEEtran}
\usepackage{cite}
\usepackage{amsmath,amssymb,amsfonts}
\usepackage{algorithmic}
\usepackage{graphicx}
\usepackage{textcomp}
\usepackage{xcolor}
\usepackage[hyphens]{url}

\def\BibTeX{{\rm B\kern-.05em{\sc i\kern-.025em b}\kern-.08em
    T\kern-.1667em\lower.7ex\hbox{E}\kern-.125emX}}

\newcommand{\todo}{\textcolor{purple}{TODO: }\textcolor{purple}}

\usepackage{braket}
\usepackage{soul}
\usepackage{booktabs}
\usepackage{adjustbox}

\usepackage{bm}
\usepackage{cite}

\usepackage[linewidth=1pt]{mdframed}

\usepackage{tikz}
\newcommand*\circled[1]{\tikz[baseline=(char.base)]{
            \node[shape=circle,draw,inner sep=0.5pt] (char) {#1};}}
\usepackage{mathtools}

\usepackage{tabularx}
\usepackage{authblk}

\newcommand\blfootnote[1]{%
  \begingroup
  \renewcommand\thefootnote{}\footnote{#1}%
  \addtocounter{footnote}{-1}%
  \endgroup
}

\title{Boosting Quantum Fidelity with an Ordered Diverse Ensemble of Clifford Canary Circuits} 

\author[1]{Gokul Subramanian Ravi$^{*}$}
\author[1]{Jonathan M. Baker}
\author[1,2]{Kaitlin N. Smith}
\author[3]{Nathan Earnest}
\author[3]{\\Ali Javadi-Abhari}
\author[1,2]{Frederic T. Chong}

\affil[1]{University of Chicago}
\affil[2]{Super.tech (a division of ColdQuanta)}
\affil[3]{IBM Quantum, IBM T.\ J.\ Watson Research Center}

\begin{document}
\maketitle
\thispagestyle{plain}
\pagestyle{plain}

%%%%%% -- PAPER CONTENT STARTS-- %%%%%%%%

\begin{abstract}

On today's noisy imperfect quantum devices, execution fidelity tends to collapse dramatically for most applications beyond a handful of qubits.
It is therefore imperative to employ novel techniques that can boost quantum fidelity in new ways. %Only then can we substantially advance quantum frontiers in the immediate future. 

This paper aims to boost quantum fidelity with Clifford canary circuits by proposing  \emph{Quancorde: \underline{Quan}tum \underline{C}anary \underline{Or}dered \underline{D}iverse \underline{E}nsembles}, a fundamentally new approach to identifying the correct outcomes of extremely low-fidelity quantum applications.
It is based on the key idea of diversity in quantum devices - variations in noise sources, make each (portion of a) device unique, and therefore, their impact on an application's  fidelity, also unique.

Quancorde utilizes Clifford canary circuits (which are classically simulable, but also resemble the target application structure and thus suffer similar structural noise impact) to order a diverse ensemble of devices or qubits/mappings approximately along the direction of increasing fidelity of the target application.
Quancorde then estimates the correlation of the ensemble-wide probabilities of each output string of the application, with the canary ensemble ordering, and uses this correlation to weight the  application's noisy probability distribution.
The correct application outcomes are expected to have higher correlation with the canary ensemble order, and thus their probabilities are boosted in this process.
%expect that produce the most similar ensemble ordering, and recognizes these as the likely correct outcomes of the target application.
%Quancorde is able to identify the top 1-10s of outputs (out of 1000s of possible output strings) for each evaluated application, among which is the correct application outcome. 

Doing so, Quancorde improves the fidelity of evaluated quantum applications by a mean of 8.9x/4.2x (wrt. different baselines) and up to a maximum of 34x.

\end{abstract}

\thispagestyle{empty}

\blfootnote{

\noindent *Correspondence: gravi@uchicago.edu. 
}

\section{Introduction}
\label{sec:intro}

%Importance of QC

Quantum computing is a revolutionary computational model that takes advantage of quantum mechanical phenomena to solve intractable problems. 
Quantum computers can potentially leverage superposition, interference, and entanglement to give significant computing advantage in chemistry~\cite{kandala2017hardware}, optimization~\cite{moll2018quantum}, machine learning~\cite{biamonte2017quantum} etc.

\begin{figure}[t]
\centering
%\fbox{
\includegraphics[width=0.98\columnwidth,trim={0.4cm 0cm 0cm 0cm},clip]{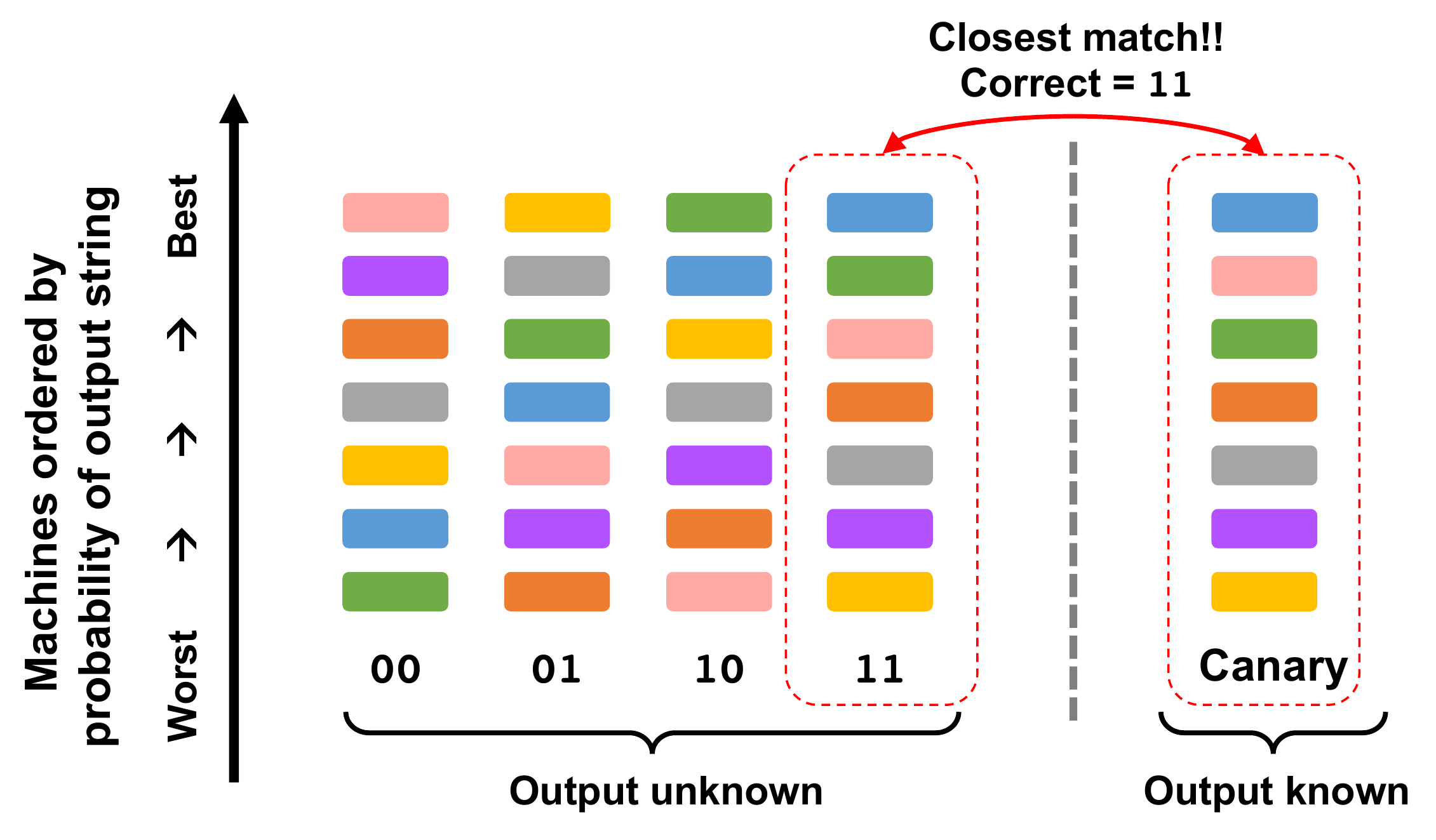}%}
\caption{
%Outputs of quantum circuits are usually unknown. 
In Quancorde, a target quantum circuit is executed on a diverse ensemble (eg., on different machines) as shown by the different colors. 
Then, for each non-negligible output bitstring produced in noisy circuit execution (from `00' - `11'), all machines are ordered by that string's occurrence probability - these are the first 4 columns.
%of occurrence - each can produce a unique ordering as shown. 
%Next, Quancorde constructs a `canary' circuit the closely resembles the target circuit structure and is efficiently classically simulable. 
Next, Quancorde runs a `canary' circuit (which resembles the target circuit structure and is classically simulable) on the ensemble to obtain a canary ordering, i.e., the machine ordering that increases the occurrence of the correct canary output. 
% (which is known classically). 
%The canary ordering is expected to be a close match to the ordering created by the correct output for the original circuit. 
Then, the output string from the original circuit that produces a machine ordering that is  closely correlated with the canary ordering is likely to be the correct outcome. This is `11' in the figure. This correlation information is then used to weight the original output distribution to boost application fidelity (Fig.\ref{fig:quancorde_overview}).
%Thus, diversity is exploited to obtain the correct output of the target circuit. 
}
\label{fig:quancorde_intro}
\end{figure}

\begin{figure*}[t]
\centering
%\fbox{
\includegraphics[width=0.9\textwidth,trim={0cm 0cm 0cm 0cm},clip]{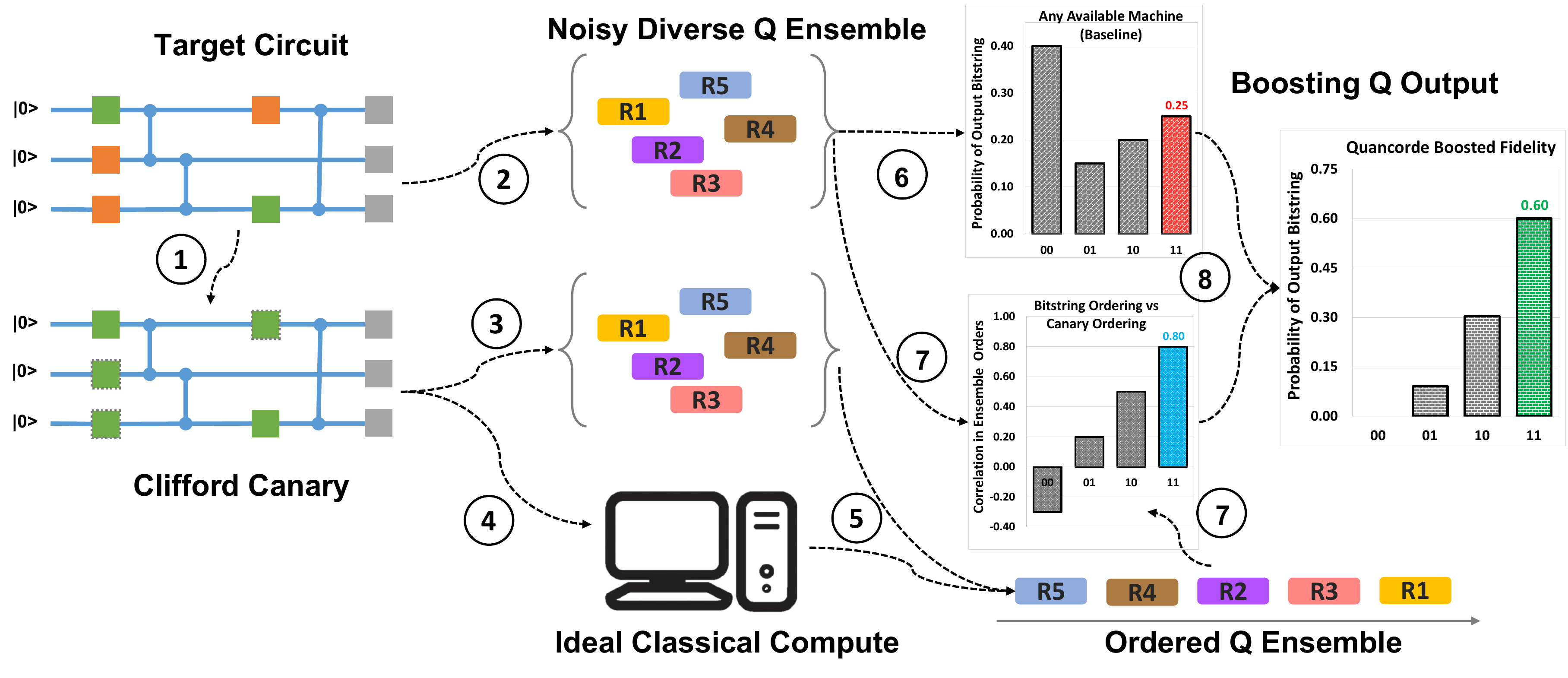}%}
\caption{Quancorde overview. (1) A canary circuit for the target application is constructed by replacing the non-Clifford gates in the target (orange) with the nearest Clifford gates (green). (2) The target circuit is executed on the diverse noisy quantum ensemble R1-R5 (e.g. different machines or qubits/mappings). (3) In parallel, the canary is also executed on the same noisy ensemble. (4) The correct output of the canary is obtained by running it ideally (noise-free) on a classical machine - possible since Clifford circuits are efficiently classically simulable. (5) Since the correct canary outcome is known, the ensemble is ordered based on the noisy execution fidelity of the canary - $R5<R4<R2<R3<R1$. (6) The noisy distribution on any machine is selected as the baseline - the correct answer `11' has low probability. (7)  Ensemble orderings are produced for the different noisy outputs (of non-negligible probability) of the original circuit and are compared with the canary ordering (like in Fig.\ref{fig:quancorde_intro}) to estimate a correlation value for each output 0 `11' has a high correlation. (8) The correlations are then used to weight the baseline distribution to produce a new  distribution in which the `11' probability is boosted to become the winner.}
\label{fig:quancorde_overview}
\end{figure*}

%NISQ era + fidelity really low
In near-term quantum computing, we expect to work  with  machines which comprise 100-1000s of imperfect qubits ~\cite{preskill2018quantum}. 
These machines suffer from high error rates in the form of state preparation and measurement (SPAM) errors, gate errors, qubit decoherence, crosstalk, etc.
While, in the near future, it is clear that we will be unable to execute large-scale quantum algorithms like Shor's Factoring~\cite{Shor_1997} and Grover Search~\cite{Grover96afast}, which would require error correction comprised of millions of qubits to create fault-tolerant quantum systems~\cite{O_Gorman_2017}, it is important to uncover some potential for quantum advantage in the near-term.

Classical computing today is able to perfectly simulate quantum systems up to around 50 qubits (with the help of supercomputers, if required)~\cite{tilly2021variational,Haner2017,Raedt2019,Boixo2018,45919}.
Thus, advancing today's quantum computing to the doorstep of quantum advantage would require us to solve quantum applications of around 50 qubits with reasonable fidelity\footnote{fidelity: the probability of finding the correct outcome or set of outcomes from the output distribution produced by a noisy quantum circuit.}.
However, on today's quantum devices, execution fidelity tends to collapse dramatically for most circuits beyond a handful of qubits.

Multiple error mitigation techniques have been explored recently.
These techniques reduce the effect of noise on circuit execution on the quantum machine.
Several promising strategies have recently emerged~\cite{czarnik2020error,Rosenberg2021,barron2020measurement,botelho2021error,wang2021error,takagi2021fundamental,temme2017error,li2017efficient,giurgica2020digital,ding2020systematic,smith2021error,murali2020software,ding2020systematic,murali2019noise,tannu2019not,viola1999dynamical,pokharel2018demonstration, souza2012robust,tannu2019mitigating,bravyi2021mitigating}.
However, the resulting execution fidelity is still minuscule for most circuits beyond 10 qubits, let alone real-world use cases. 
Thus, while we must continue to innovate across the hardware and software stack to improve the fidelity we can get out of any given quantum device, it is also imperative to ponder entirely different techniques that are able to boost quantum application fidelity in new ways, beyond the fundamental limitations of today's noisy quantum devices. Only then can we advance quantum frontiers in the near future.

%Go to contributions

% Talk about diversity
This work proposes one such fundamentally new approach to identifying the correct outcomes of extremely low  fidelity quantum applications.
It is based on the key idea of diversity in quantum devices - no two quantum devices are identical.
Today's device errors stem from multiple noise sources such as the imperfect classical control of the device, thermal fluctuations, destructive qubit coupling, imperfect insulation of the qubits, quasi-particles, and other external stimuli~\cite{muller2019towards,martinis2005decoherence,burnett2019decoherence,schlor2019correlating,Klimov_2018}.
Because current fabrication techniques lack the precision to make homogeneous batches of quantum devices, the noise properties are distinct for each device.
Furthermore, the dynamic nature of quantum systems causes these noise sources to suffer from spatial and temporal variation. 
Thus, each (portion of a) device is unique, and its impact on an application's execution / fidelity, is also unique.

Simply put, a device which is known to be `better' for the target application would produce higher application fidelity (compared to other devices).
In other words, it would produce a higher probability of occurrence of the correct application outcomes in its output distribution.
%(even if this probability was relatively low).
Extending this notion further, every output string that an application might produce in noisy execution, will have a unique ordering of devices that increase the string's probability of occurrence from low to high.
If one of these orderings is somehow known to be the `correct' device ordering that improves the fidelity of the application, then the bitstring that produces this device order would likely be among the correct outcomes of the application.
Of course, knowing this correct ordering is not trivial, but in this work we show that this is very feasible.
Note: this diversity could also be different qubit mappings within a single device.

%Proposal and Fig.1 - come at it from diversity point of view, then say that we have this canary method and why this works
We propose \emph{Quancorde: Boosting fidelity with \underline{Quan}tum \underline{C}anary \underline{Or}dered \underline{D}iverse \underline{E}nsembles}.
Quancorde exploits diversity and ordering, as expressed above, to identify the correct outcome of a target application.
It is introduced in Fig.\ref{fig:quancorde_intro}.
In Quancorde, a target quantum circuit is executed on a diverse ensemble of resources, (eg., on different machines) as shown by the different colored blocks in the figure. 
Then, for each output bitstring of the circuit (of non-negligible occurrence), machines are ordered by the string's probability of occurrence.
Each bitstring can produce a unique ordering as shown for 4 different output strings (`00',`01',`10' and `11') for a 2-qubit circuit in the figure. 
Now, the critical step is to identify the (nearly) correct ensemble ordering that improves the application fidelity.

To do so, Quancorde constructs a quantum `canary' circuit (inspired in name by prior classical canary circuits~\cite{Ernst:2003}). 
The canary circuit is important in two ways:
\circled{1}\ First, the canary circuit is made up of only Clifford gates and is constructed by replacing all non-Clifford gates in the target circuit with the `nearest' Clifford gates. 
A circuit made up of only Clifford gates is efficiently classically simulable~\cite{gottesman1998heisenberg},  and thus the correct output of the canary circuit is obtained via ideal classical simulation.
\circled{2}\ Second, the canary circuit maintains the exact device-mapped circuit structure of the original circuit - it has the same circuit critical depth/paths, the same 2-qubit CNOT gates, and the same measurement bits.
Thus, it suffers from the same noise sources, and their relative impact on its circuit fidelity is similar to that on the original circuit, even if the quantum states explored by the two circuits are different.
Next, Quancorde runs this canary circuit on the ensemble to obtain a canary ordering, i.e. the ensemble ordering for the correct canary output (which is known from \circled{1}) - this is shown to the right of the figure.
The canary ordering is expected to be a close match to the ordering created by the correct output for the original circuit (as discussed in \circled{2}). 
Next, the ensemble order produced by each noisy output string of the application is correlated against the canary ordering.
A correlation distribution is obtained in which the correct outcome of the application is likely to have a strong correlation to the canary order (again, from \circled{2}).
In the figure, this corresponds to the `11' string, whose ordering is almost identical to the canary ordering.
Finally, this correlation distribution is used to weight the original application noisy distribution.
Since the likely correct outcomes are weighted higher, this tremendously boosts application fidelity.
%Finally, the output string from the original circuit that produces an ensemble ordering that is the closest match to the canary ordering is deemed the correct output. 
%Thus, `11'  is identified as the correct output of the target 2-qubit application.
More the diversity of the ensemble, greater is the uniqueness of the ensemble order, creating a more well defined correlation distribution,  thus, leading to a larger boost to application fidelity via the Quancorde approach.
A step-by-step breakdown of Quancorde is shown in Fig.\ref{fig:quancorde_overview}.
%and is discussed in its caption.

\textbf{Overall, we make the following contributions:}

%\begin{enumerate}
\circled{1}\ We propose a fundamentally new approach to identifying the correct outcomes of extremely low fidelity quantum applications by exploiting quantum device diversity.

\circled{2}\  We propose Clifford canary circuits to order a diverse ensemble of devices or qubits/mappings along the direction of increasing fidelity of the target application.

\circled{3}\  We then identify the correlation of the application outputs' ensemble orderings with the canary order, and use this to weight the application output distribution, resulting in boosted application fidelity.

\circled{4}\  The above ideas are presented as Quancorde, which improves the fidelity of our evaluated quantum applications (with baseline fidelity as low as 0.1\%) by a mean of 8.9x/4.2x (wrt. different baselines) and a maximum of 34x. 

\circled{5}\ Importantly, Quancorde can surpass fundamental limitations of any particular noisy quantum device and thus improve application fidelity beyond the capability of any single device.
%\end{enumerate}

\iffalse
\emph{Quancorde is especially useful when diversity is significant, and applications' ideal outcomes are hard to produce - both of which are intuitive expectations in the near quantum computing future. Quancorde can have a revolutionary impact on the way noisy quantum resources are effectively leveraged to boost the fidelity of real-world quantum use cases.}
\fi

\section{Background and Motivation}
\label{sec:bm}

\begin{figure}[t]
\centering
%\fbox{
\includegraphics[width=0.98\columnwidth,trim={0.5cm 0.5cm 1cm 0cm},clip]{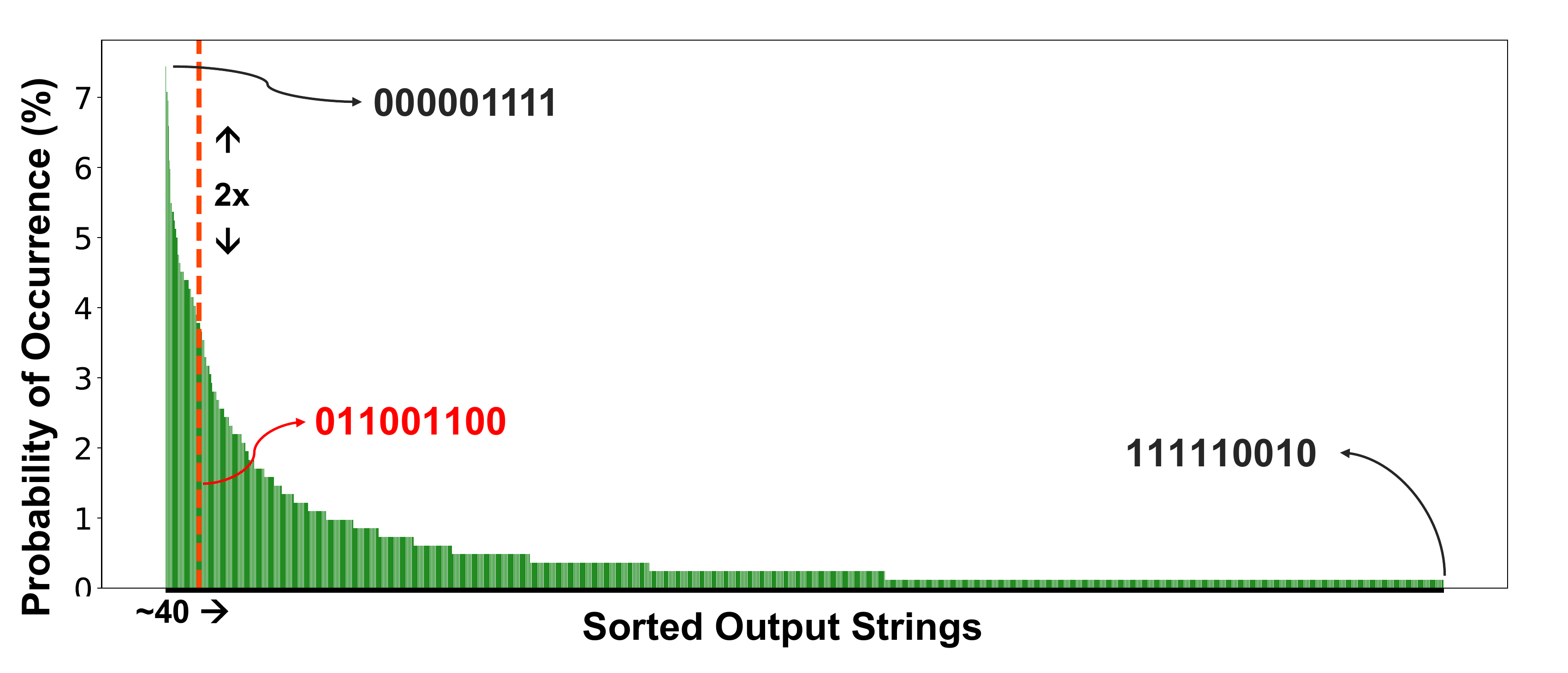}%}
\caption{Sorted probability of occurrence for output trings produced by a 9-bit circuit executed on IBMQ Montreal. The correct outcome is indicated in red. 
%The probability of the correct outcome is very low at 4\% and is also 2x lower than the probability of the highest occurring outcome. %Further, there are nearly 40 bitstrings with probabilities greater than the correct outcome.
}
\label{fig:quancorde_motive_fid}
\end{figure}

\subsection{Fidelity in the noisy quantum era}

\label{b_NISQ}
Today's quantum devices are error-prone and up to around 100 qubits in size~\cite{preskill2018quantum}. 
%These devices are extremely sensitive to external influences and require precise control, and as a result, some of the biggest challenges that limit scalability include limited coherence, gate errors, readout errors, and connectivity. %TODO CAMERA
Multiple error mitigation strategies have been proposed to correct different forms of quantum errors.
These include, but are not limited to, noise-aware compilation~\cite{murali2019noise,tannu2019not}, scheduling for crosstalk~\cite{murali2020software,ding2020systematic}, 1Q gate scheduling in idle windows~\cite{smith2021error}, dynamical decoupling~\cite{viola1999dynamical,pokharel2018demonstration, souza2012robust}, zero-noise extrapolation~\cite{temme2017error,li2017efficient,giurgica2020digital}, readout error mitigation~\cite{tannu2019mitigating,bravyi2021mitigating}, exploiting quantum reversibility~\cite{patel2021qraft,smith2021error} and many more~\cite{czarnik2020error,Rosenberg2021,barron2020measurement,botelho2021error,wang2021error,takagi2021fundamental,temme2017error,li2017efficient,giurgica2020digital}.
In addition, some of these can be used in conjunction to achieve better fidelity~\cite{ravi2021vaqem}.
However, the resulting fidelity is still insufficient for most circuits beyond 10 qubits, with the correct circuit outcome(s) often not being among the most dominant in the output distribution.

Fig.\ref{fig:quancorde_motive_fid} shows the sorted probability of the output bitstrings obtained from a 9-qubit circuit executed on IBM Quantum Montreal with the highest noise-aware compiler optimizations. 
The correct outcome is shown in red. 
Not only is the probability of the correct output (i.e., fidelity) very low at 4\%, it is also 2x lower than the probability of the highest occurring output. 
Furthermore, the output distribution has nearly 40 bitstrings with occurrence probabilities higher than that of the correct outcome. 
It is evident that for circuits such as this, and those of greater complexity, today's machines will be very limited in the fidelity they achieve.
While error mitigation techniques improve fidelity and will continue to do so as more novel solutions are proposed, it is clear that noisy quantum machines will be fundamentally limited, and we  require entirely novel approaches to identify the correct outcomes of applications of real-world criticality.
%of hard-to-execute circuits, which is critically needed for real-world applicability.
In this work, we turn to diversity.

\begin{figure}[t]
\centering
%\fbox{
\includegraphics[width=0.9\columnwidth,trim={0.1cm 0cm 0.4cm 0cm},clip]{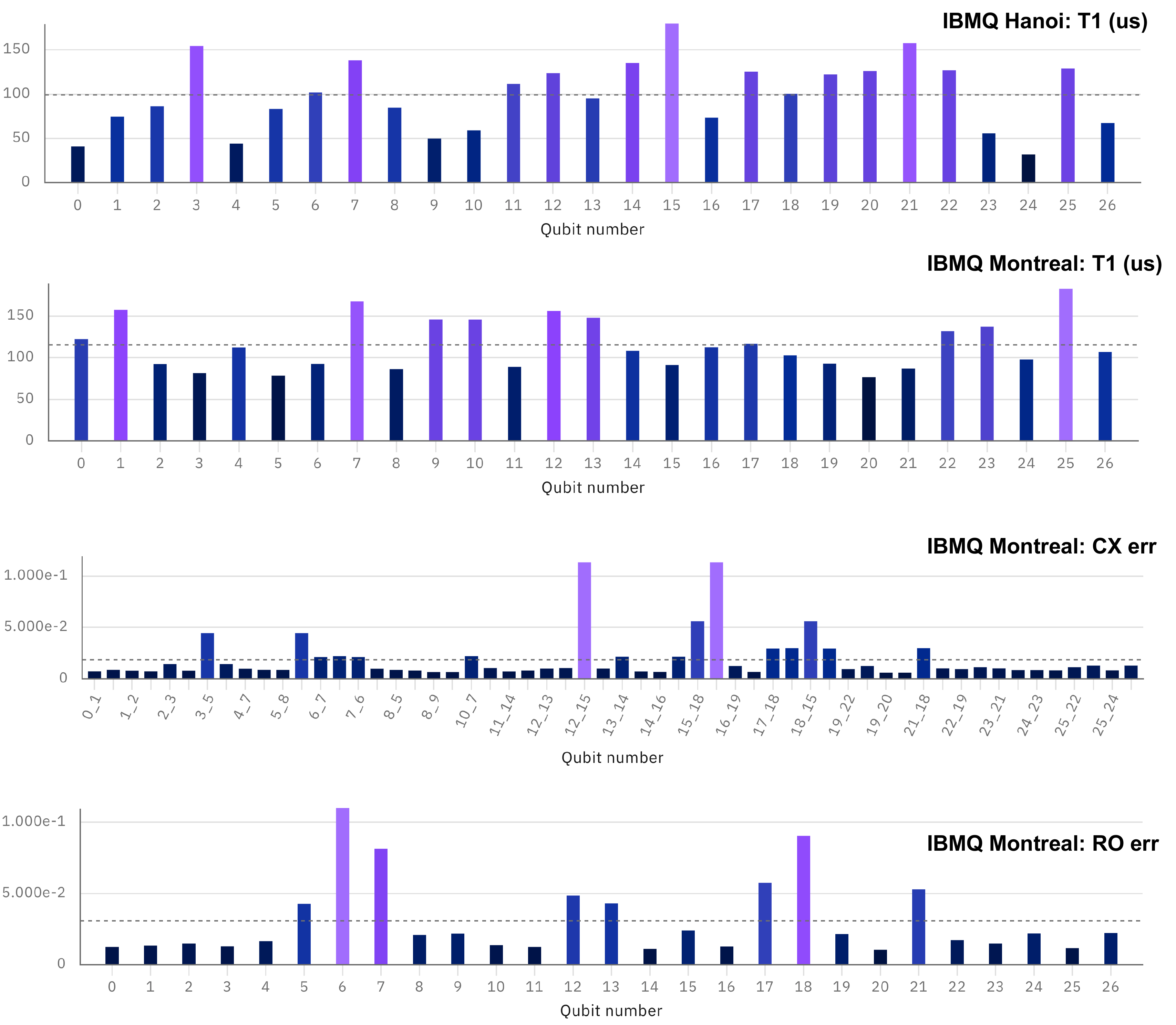}%}
\caption{Diversity in IBM Quantum machine characteristics, obtained from ~\cite{IBMQE} on 04/17/2022. The top 2 figures show T1 times across the qubits on IBM Quantum Hanoi and Montreal respectively. 
%Clearly there is considerable variation across the machines and within each machine. %TODO CAMERA
The bottom 2 figures show CNOT and Readout errors for each qubit on Montreal. 
%Again, the diversity in error rates is evident. Circuits executed on different machines or on different qubits, will experience different noise and produce different output distributions.  %TODO CAMERA
}
\label{fig:quancorde_motive_div}
\end{figure}

\subsection{Diversity in Quantum Devices}

Even if QCs are manufactured in a highly controlled setting, unavoidable variation results in intrinsic properties that impact performance. 
%This variation between and within devices becomes especially apparent when examining error rates. 
Fig.\ref{fig:quancorde_motive_div} shows data on IBM Quantum machine characteristics obtained from ~\cite{IBMQE} on 04/17/2022.
%They are indicative of the substantial diversity that is present across different qubits within a quantum device, and the diversity across different devices.
The top two figures show T1 coherence times across the 27 qubits on IBM Quantum Hanoi and Montreal, respectively. 
These coherence times are indicative of the circuit durations at which amplitude damping occurs.
%A lower T1 time results in a $\ket{1}$ state falling to the $\ket{0}$ state faster.
In the data shown, the average coherence time of Hanoi is lower than that of Montreal, indicating a higher likelihood of damping. 
Furthermore, some qubits are much worse than others.
On Montreal, a 4-qubit circuit mapped to qubits (2, 3, 5, 6) would decohere faster than one mapped to (9, 10, 12, 13).
Thus, clearly the circuit fidelity (especially for deeper circuits) could be very different across these two mappings and devices.
The two figures at the bottom show CNOT errors (error on 2-qubit CNOT gates) and Readout errors (error on qubit measurements) for qubits on Montreal. 
Again, the diversity in error rates is evident. 
A circuit with a CNOT gate between qubits (12, 15) would likely have lower fidelity than a circuit with the same CNOT gate instead between qubits (14, 16).
Similar impact of readout errors can be inferred.

From the above discussion, it is intuitive that the same application executed on different machines or on different qubits/mappings within a machine, will experience different noise characteristics and therefore produce different output probability distributions.
%Further, considering the heterogeneity in qubit connectivity within devices and across devices (for examples, see \cite{IBMQE}), choice of different devices or qubits/mappings can create very different post device mapped circuit structures. %TODO CAMERA
%For example, a deeper circuit with more CNOT gates would likely have lower fidelity.
%It is also worth noting that these characteristics can change substantially every time the devices are calibrated (roughly once a day for IBM Quantum machines) and also drift between calibrations~\cite{Proctor_2020}.
Finally, it is important to recognize that for circuits of even reasonable complexity, the fidelity impact cannot be directly inferred simply by observing error rates and decoherence times.
We discuss this further in Fig.\ref{fig:quancorde_cliff_2} and Section \ref{e:quant_can}, but this is intuitive, since if this were possible, then noise-aware quantum simulators would be perfect at mimicking real device fidelity (which they are not).

\begin{figure}[t]
\centering
%\fbox{
\includegraphics[width=0.98\columnwidth,trim={0cm 0cm 0cm 0cm},clip]{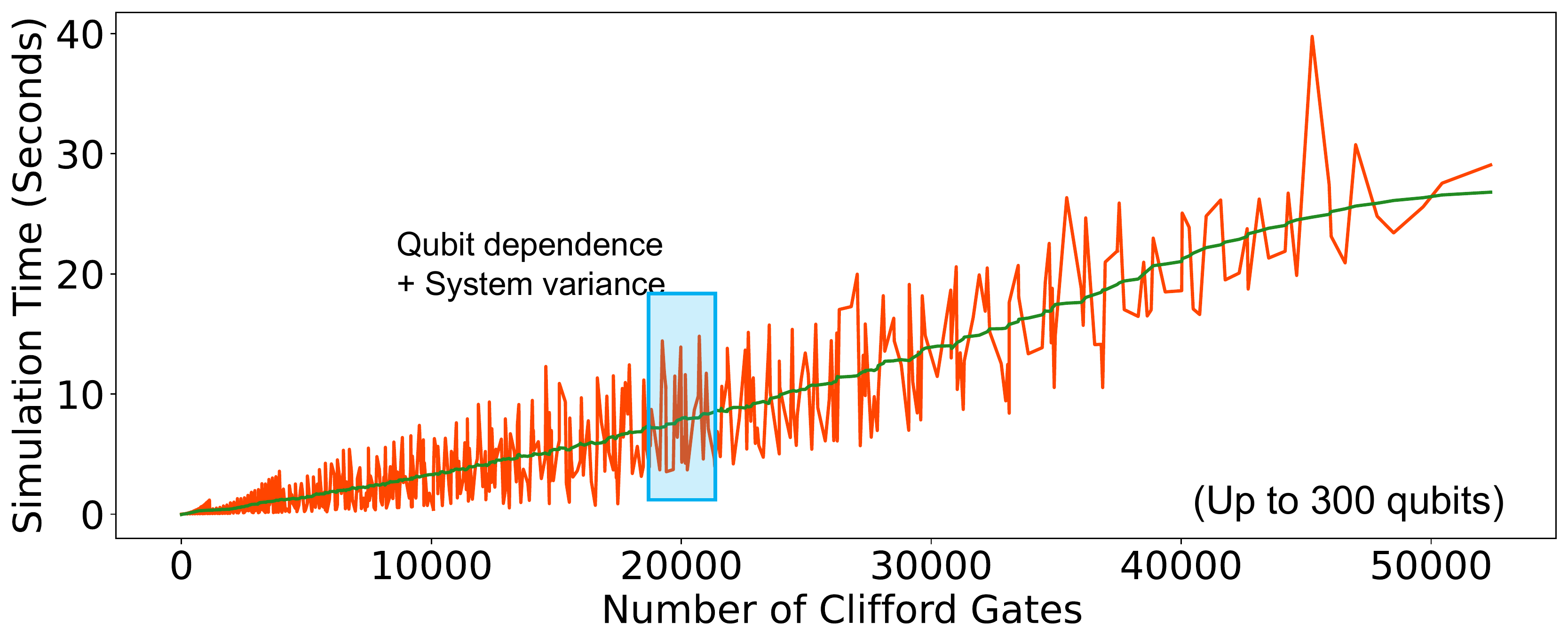}
%}
\caption{Simulating Clifford circuits on the IBMQ Stabilizer Simulator~\cite{IBMQE} run on a laptop. The linear scaling with number of gates is clearly visible. Circuits with 300 qubits / 50,000 gates can be simulated in 30 seconds.}
\label{fig:eval_cliff_scaling}
\end{figure}

\begin{figure*}[t] %moved up
\centering
%\fbox{
\includegraphics[width=\textwidth,trim={0.2cm 0.5cm 0cm 0cm},clip]{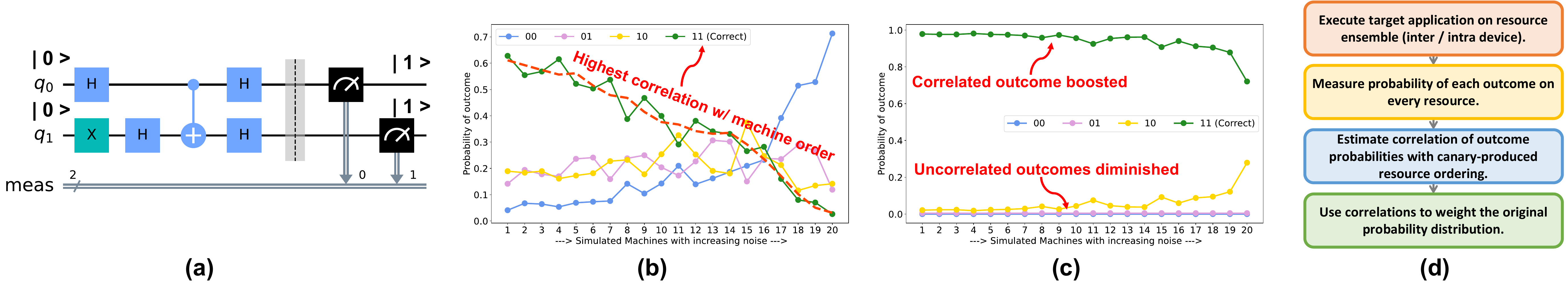}%}
\caption{Leveraging an ordered diverse ensemble to boost application fidelity. (a) 2-qubit circuit with a `11' correct output. 
(b) The 2-qubit circuit is executed on 20 simulated machines with increasing noise, and the output bitstring probabilities are plotted. The probability of the correct bitstring `11' decreases in high positive correlation with decreasing machine quality, whereas other bitstrings have low / negative correlation. 
%On the other hand, `00' increases as machine quality worsens (negative correlation), while `01' and `10' are poorly correlated. 
(c) The correlations are used to weight the output probabilities. The correct bitstring is boosted due to high correlation, while others are diminished. Thus, it stands out as the clear estimated outcome. (d) Flow chart of the Quancorde proposal assuming a given canary-ordered ensemble.}
\label{fig:quancorde_proposal_qc}
\end{figure*}

\subsection{{Clifford Circuits}}
\label{bm_cliff}
Classical simulation of quantum problems usually requires exponential resources (otherwise the need for quantum computers is obviated).
Even using high-performance supercomputers, the simulation is restricted to around 50 qubits~\cite{tilly2021variational,Haner2017,Raedt2019,Boixo2018,45919}.
An exception to the above is the classical simulation of the Clifford space which is a subset of the total quantum Hilbert space.
Circuits made up of only Clifford operations can be exactly simulated in polynomial time (almost linear in the number of gates and qubits) as stated by the Gottesman-Knill theorem~\cite{gottesman1998heisenberg}.

We quantitatively showcase this in Fig.\ref{fig:eval_cliff_scaling}.
The figure shows simulation time for purely Clifford circuits of up to 50,000 gates and 300 qubits.
These are simulated on the IBM Stabilizer Simulator~\cite{IBMQE} run on a laptop computer.
Clearly the simulation time is seen to scale fairly linearly in the number of gates (with some variation stemming from number of qubits in each circuit as well as computer load).
Further, even the largest circuits are simulated in roughly 30 seconds.
In contrast, simulating non-Clifford circuits of this size is impossible even with the most powerful supercomputers.
%would require supercomputers even for 40 qubits with 1000 gates.
%The Gottesman-Knill theorem states that \emph{``Any quantum computer performing only: a) Clifford group gates, b) measurements of Pauli group operators, and c) Clifford group operations conditioned on classical bits, which may be the results of earlier measurements, can be perfectly simulated in polynomial time on a probabilistic classical computer"}~\cite{gottesman1998heisenberg}.

While the Clifford group operations do not provide a universal set of quantum gates, the benefits of classical simulation of Clifford circuits is clearly evident above and can be leveraged in interesting ways.
Some prior applications are discussed in Section \ref{RW}.
In this work, Clifford circuits are leveraged to construct canary circuits that closely resemble the structure of the target application. 
By being classically simulable, the correct outcomes of the canary circuits can be estimated on a classical computer in an efficient manner.
This is then utilized towards ordering the diverse ensemble. 

\iffalse
We note that it is possible to have a small number of non-Clifford gates in circuits and still be efficiently classically simulable, we discuss this in Section \ref{FW}.
\fi

\section{Leveraging ordered ensembles}
\label{p:ens}

In this section, we discuss how ensemble ordering enables the boosting of noisy application fidelity.
%identification of the correct outcome, or, at the very least, a set of outputs which is likely to contain the correct outcome, of a  target quantum application.
This section assumes that an ensemble order that is well correlated to the fidelity of the target application is known; the practical implementation of this, with canary circuits, is discussed in Section \ref{p:can}.

%\subsection{Illustrative example}

Fig.\ref{fig:quancorde_proposal_qc}.a shows a 2-qubit circuit that exploits phase kickback (integral to applications like Bernstein-Vazirani~\cite{bv}) to produce `11' as the correct  outcome.
When executing on real machines, such a circuit would suffer from noise and 
%decoherence, 1-qubit and 2-qubit gate errors, readout errors, and crosstalk, with the impact of these error forms increasing with circuit complexity. 
thus, the output distribution produced is a set of many bitstrings, which in this case are `00', `01', `10' and `11'.
Our goal is to identify that `11' is the correct outcome.
Note that in the presence of significant noise, `11' is not necessarily the most probable output (similar to Fig.\ref{fig:quancorde_motive_fid}), so simply capturing the output with highest probability is often insufficient.

In this experiment, we execute this 2-qubit circuit on a diverse ensemble of 20 diverse simulated `machines' - 20 simulations with varying error characteristics.
For this experiment, we assume that the ensemble ordering that is well correlated with the circuit's fidelity is known (i.e., we have an intuition for the ordering of machines that improves the fidelity of the application, even if we don't know the correct outcome of the application). 
%For trivial circuits like Fig.\ref{fig:quancorde_proposal_qc}, this ensemble ordering can be naively derived using the product of different noise components experienced by the circuit - decoherence, gate errors, RO errors etc.
%Since this experiment is simulating machines with varying error, we directly infer the ordering from the error quantities.
All errors increase uniformly from machines 1 to 20, so we assume that the reverse ordering will be well correlated with application fidelity. 

Now, recall that we still do not know the correct outcome of our target circuit. 
Next, we plot the probability of occurrence of each output string against the error-ordered machine ensemble.
This is shown in Fig.\ref{fig:quancorde_proposal_qc}.b.
Clearly the probability of the `11' outcome (green) falls almost monotonically as we go from best (1) to worst (20) machines, showing high (nearly perfect) positive correlation with the `correct' machine order.
On the other hand, the `00' outcome  probability (blue) increases as machine quality worsens (negative correlation) and the `10' and `01' outcomes (yellow / plum) have no discernible correlation.

Next, we use these correlations to weight the original application's output probability distribution (this weighting is similar to Bayesian Reconstruction~\cite{jigsaw}).
Outputs with high positive correlation are boosted, those with low positive correlation are diminished, and those with negative correlation are simply removed.
This can be done for the application's output distribution on any machine or every machine or a known fairly good machine.
We show this for every machine in  Fig.\ref{fig:quancorde_proposal_qc}.c.
Clearly, the `11' outcome probability is tremendously improved due to its high positive correlation. 
Now, from this figure, it is easy to identify that `11' is the correct application outcome since it clearly stands apart from other outputs. 
The resulting fidelity boost is significant for any machine in the ensemble, and clearly showcases the capability of Quancorde.
A flow chart of the Quancorde proposal, generalizing the steps to exploit a diverse ensemble is shown in Fig.\ref{fig:quancorde_proposal_qc}.d and was also illustrated in Fig.\ref{fig:quancorde_overview}.

Greater diversity and size of the ensemble are likely to imply more comprehensible identification of the correct outcome or a smaller subset of outcomes.
Note that the ensemble does not have to be unique devices, it could also be different sets of qubits or different mapping of qubits with a single device or both - thus ensemble size can be increased. Diversity focused extensions are discussed in Section \ref{FW}.

%Thus, we are able to identify `11' as the correct outcome for this circuit, since its probability of occurrence increases in high correlation with improving machine quality.
%Using ordered ensembles to identify the correct outcome is illustrated in Fig.\ref{fig:quancorde_overview} and described in the caption bullets (2) and (6).

%\subsection{Practical use beyond the correct outcome}
%\label{sec:practical}

\textbf{\emph{Identify the correct outcome with just correlations?:}}
In Fig.\ref{fig:quancorde_proposal_qc}.b, we estimated that the `11' output has the highest correlation with the `correct' machine order.
At that point, we could have directly identified `11' to be the correct outcome without having to weight the original probability distribution, simply by its highest correlation.
However, this can be insufficient for very low fidelity applications, since it it is possible that Quancorde does not actually identify the correct outcome as the one with the highest correlation.
In that scenario, one could instead consider using a subset of the top correlated outcomes,
%this is  useful for oracle based algorithms such as Bernstein-Vazirani~\cite{bv}, Deutsch-Jozsa~\cite{dj}, and even Shor's factorization~\cite{Shor_1997} and search algorithms~\cite{grover1996fast}, as well as variatialtional algorithms~\cite{farhi2014quantum,peruzzo2014variational}.
but then identifying the size of the subset becomes challenging.
Further, there are more complications for applications with multiple correct outcomes, especially when they have unequal weights --- identifying multiple outcomes with appropriate weights from the subset is not intuitive.
These challenges are alleviated in our proposed approach of weighting the original probability distribution with the correlation distribution. 

\iffalse
Although perfectly identifying the correct outcome would be the ideal goal, there is considerable practical benefit to identifying even a reasonable subset of outcomes that are likely to contain the correct one.
In fact, reducing the search space to one that is classically efficient to explore is the quantum advantage that many of these algorithms produce.
Similarly, variational algorithms benefit tremendously if a set of good initialization states or a bounded region of the variational space is identified.
In this work, we limit ourselves to finding a small subset of outputs that are likely to contain the correct outcome, but we discuss potential extensions in Section \ref{FW}.
In Section \ref{sec:method}, we define novel metrics to evaluate the usefulness of this approach and also derive the traditional fidelity metric from these for meaningful comparisons.
\fi

\section{Canaries for ensemble ordering}
\label{p:can}

\iffalse
\begin{figure*}[t]
\centering
%\fbox{
\includegraphics[width=\textwidth,trim={0cm 0cm 0cm 0cm},clip]{figures/QC_Cliff_full_1.pdf}%}
\caption{(a) Generating Clifford canary circuits. The top circuit shows a target 4-qubit circuit after mapping to an IBM Quantum device. The bottom circuit shows a Clifford canary generated for the target circuit. 
(b) High correlation in ensemble-wide fidelity between a 10q adder (orange) and its canary (green), across an ensemble of 50  mappings on IBM Quantum Montreal. Though the output produced by the circuits are different (and at low fidelity), the two trends are fairly similar across the ensemble. 
(c) A 4-qubit QAOA circuit's fidelity across an ensemble of 3 mappings each on 7 IBM Quantum machines is shown (orange). Fidelity of the corresponding canary is also shown (green) and is seen to achieve near-perfect correlation with the original. Also shown is the fidelity predicted by the Qiskit simulation noise model (blue), that uses the day's machine calibration data. 
% It is generated by `rounding' the non-Clifford RZ gates to the nearest Clifford gates. Two instances are highlighted with orange/green circles. RZ($-15\pi/16$) is rounded to RZ($\pi$). Similarly RZ($3\pi/16$) is rounded to RZ($0$). %TODO CAMERA
}
\label{fig:quancorde_proposal_clifford}
\end{figure*}
\fi

\begin{figure}[t]
\centering
%\fbox{
\includegraphics[width=0.95\columnwidth,trim={0.4cm 0.5cm 0cm 0cm},clip]{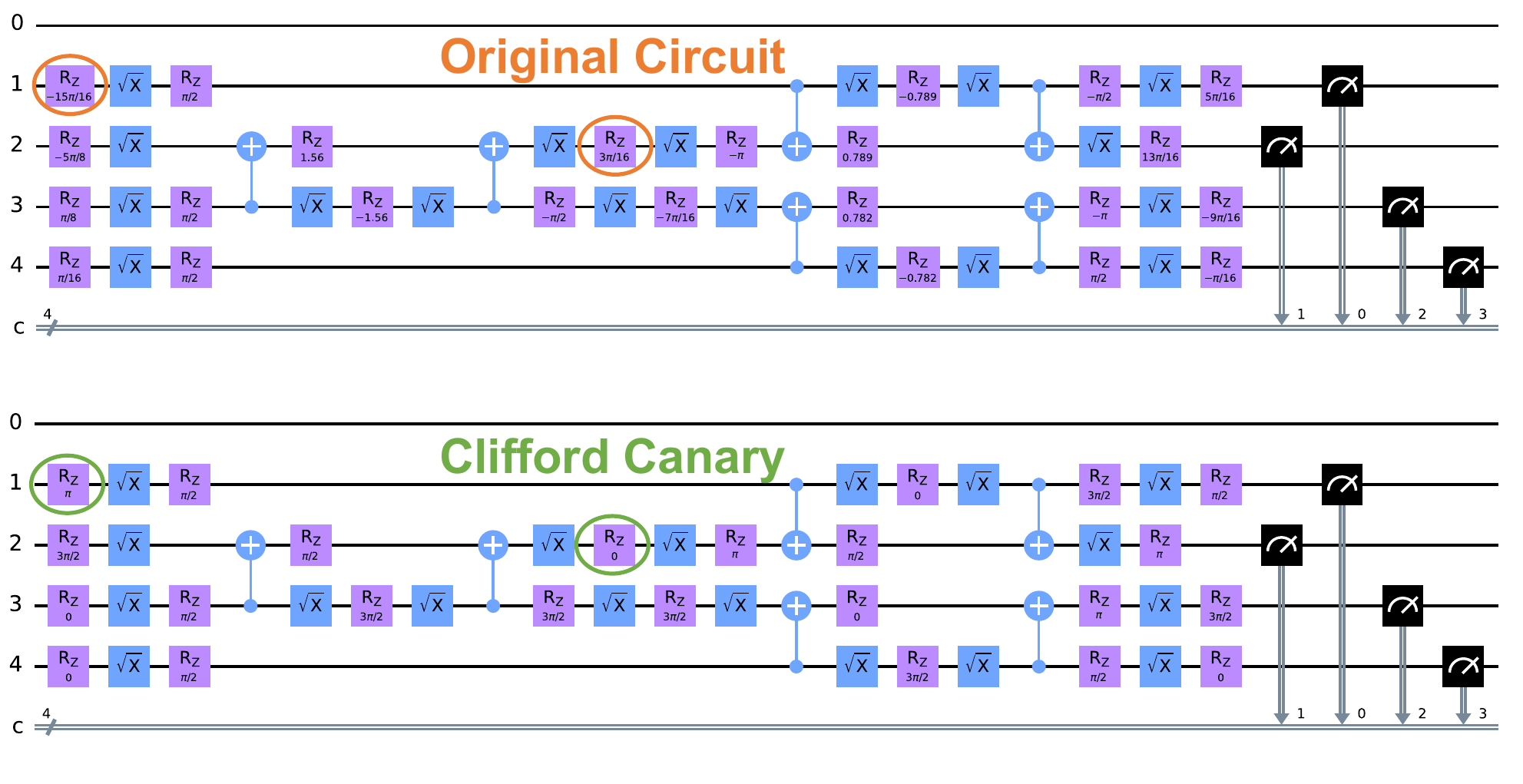}%}
\caption{Generating Clifford canary circuits. The top circuit shows a target 4-qubit circuit after mapping to an IBM Quantum device. The bottom circuit shows a Clifford canary generated for the target circuit. 
% It is generated by `rounding' the non-Clifford RZ gates to the nearest Clifford gates. Two instances are highlighted with orange/green circles. RZ($-15\pi/16$) is rounded to RZ($\pi$). Similarly RZ($3\pi/16$) is rounded to RZ($0$). %TODO CAMERA
}
\label{fig:quancorde_proposal_clifford}
\end{figure}

In Section \ref{p:ens}, we discussed how knowing an ensemble ordering that is well correlated with the fidelity of the target application can help identify the correct application outcome.
While a `good' ensemble ordering is fairly trivial to estimate for very simple circuits (roughly proportional to a product of different noise impacts), this is not the case for more complex circuits, for which fidelity is dependent on more complex interactions among the qubits and among the different noise sources.
Thus, to help find an appropriate ensemble ordering we propose `canary' circuits, which are designed with two insights.
%`Canary' circuits have been popular in the classical computing world, for instance, to understand the impact of timing errors~\cite{Ernst:2003}.

\circled{1}\ First, we observe that the impact of noise on the fidelity of a circuit is closely tied to the device-mapped structure of the circuit: how many 2-qubit CNOT gates are present and on which qubits they are placed; which qubits in the circuit are involved in critically long, potentially decohering paths; which qubits contribute to worse readout errors, etc.
Thus, a canary circuit that closely mimics the device-mapped structure of the original circuit,  plus has as much overlap as possible with the original output state, could be useful for understanding the ensemble-wide impact of noise on the original circuit.
%The intuitive reason why this is true, is because the noise from the structure of the circuit is observed to have a substantially high effect on ensemble ordering. 
%This means that knowing key information about the circuit structure, plus having as much overlap as possible with the correct output state is an effective reference circuit for noise-aware optimization.
%The fidelity ordering is significantly impacted by CX errors, measurement errors, circuit structure - all of which remain the same between the original and the canary circuits. 
But for this to be feasible, the correct outcome of the canary circuit should be known.

\circled{2}\ Thus, second, we construct these canary circuits with only Clifford gates, by replacing all non-Clifford gates in the target circuit with the `nearest' Clifford gates. 
A purely Clifford circuit is efficiently classically simulable (Section \ref{bm_cliff}), thus the correct output of the canary circuit is obtained via ideal classical simulation.
Note that the correct output of the canary is almost always different from that of the original circuit (however, choosing the nearest Cliffords helps maintain  state overlap to the best extent possible).
But importantly, it still maintains the structure of the original circuit since only specific 1-qubit gates in the circuit have to be replaced by Clifford gates.
Since the correct canary output is known, the ensemble can be ordered by the improving fidelity order of the canary circuit on the ensemble. 
This is estimated by running the canary circuit over the noisy ensemble and evaluating the probability of the correct canary output that they each produce.
Then, as discussed above, this ordering is considered to be well suited to the original circuit (and we empirically show this to be the case). 
This ordering is then utilized as described in Section \ref{p:ens} to deduce the correct outcome of the target circuit.
Construction and use of canaries is also illustrated in Fig.\ref{fig:quancorde_overview}.
%and described in caption bullets (1) and (3)-(5).

\subsection{Designing clifford canary circuits}
Fig.\ref{fig:quancorde_proposal_clifford} shows a circuit example of Clifford canary generation.
The top circuit shows a target 4-qubit circuit after mapping to a IBM Quantum device. 
On these devices, the basis gates are {CX, ID, RZ, SX, X}~\cite{IBMQE}.
These basis gates are such the only non-Clifford gates are rotational gates about the Z-axis (RZ) with angles that are not multiples of $\pi/2$. 
Any number of such gates can be present in a circuit. 
The bottom circuit shows a Clifford canary generated for the target circuit. 
It is generated by `rounding' the non-Clifford RZ gates to the nearest Clifford gates (i.e., to nearest multiple of $\pi/2$). 
%For example, RZ($-15\pi/16$) = RZ($17\pi/16$) in the upper circuit (orange circle) is rounded to RZ($\pi$) (green circle).
%Similarly, RZ($3\pi/16$) is rounded to RZ($0$), which can then be trivially removed from the circuit. 
%Many other conversions can be observed in the figure. 
Since all gates in the canary are Clifford, it can be efficiently classically simulated (Section \ref{bm_cliff}), and thus the canary correct outcome is known.
Also, it is evident from the figure that the circuit structure is maintained in the canary circuit.
%This means that the noise sources on both the circuits are exactly the same, even if the exact impact of this noise on the circuit output could, in theory, be different.
As discussed earlier, the structural similarity leads to similar fidelity trends across the ensemble for both the original and the canary.
%we observe that the impact on the circuit fidelity is, in practice, very similar, even when the outputs are somewhat different between the original and the canary.

\begin{figure}[t]
\centering
\includegraphics[width=0.95\columnwidth,trim={0cm 0cm 0cm 0cm},clip]{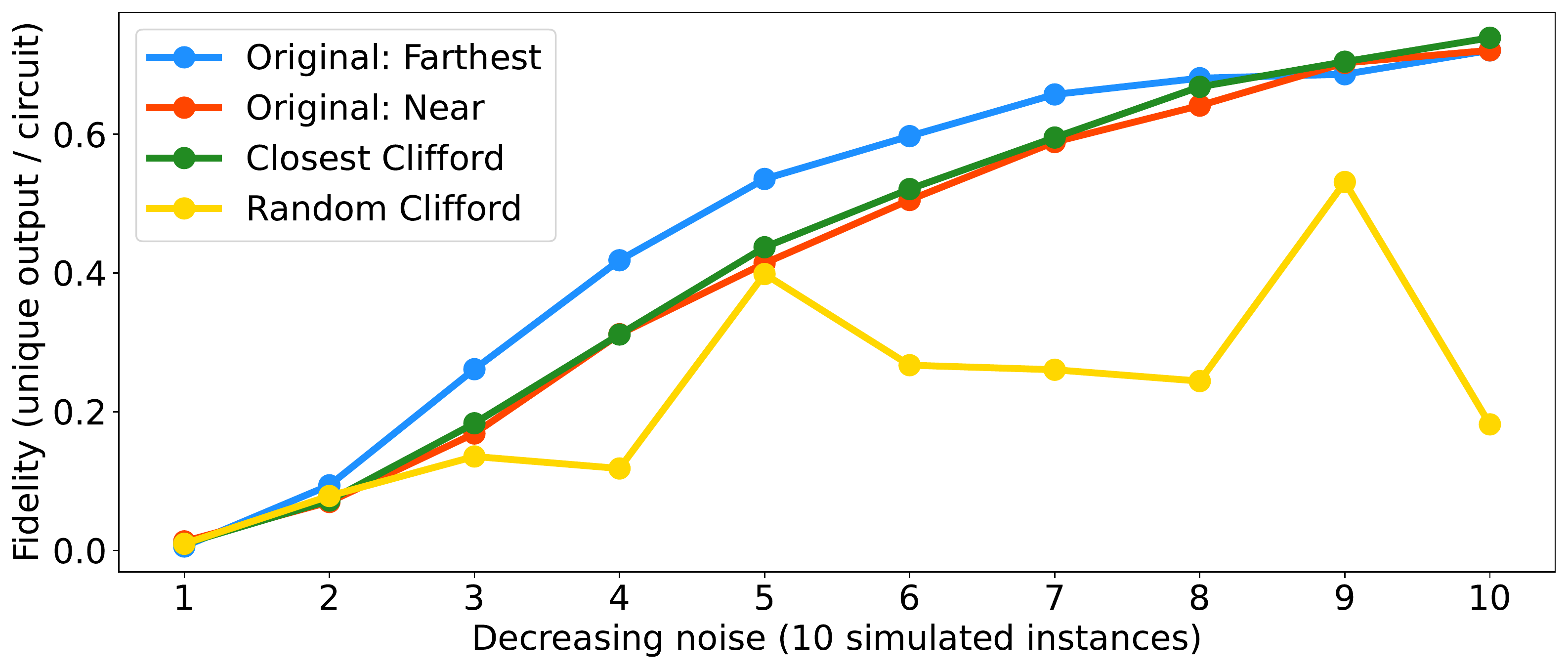}
\caption{ Fidelity over a noisy ensemble for a 10-qubit circuit. Figure shows - a)  `Original Farthest': circuit with all 1-qubit gates midway between Cliffords, b) `Original Near':  same circuit structure, with all 1-qubit gates as near-Cliffords, c) `Closest Clifford': with all 1-qubit gates as the closest Cliffords to the prior, d) `Random Clifford':  with all 1-qubits as random Cliffords. 
%This shows the accurate fidelity trend mimicking by the Closest Clifford even when the Original circuit is Farthest from it (but not so for a Random Clifford).
%The `Closest Clifford' closely mimics the fidelity trend of the Originals, but the `Random Clifford' does not. Further, the `Closest Clifford' is marginally more closely matched to `Original Near' than `Original Farthest'.
}
\label{fig:quancorde_cliff_corr_near_far}
\end{figure}

We quantify these observations in Fig.\ref{fig:quancorde_cliff_corr_near_far}.
All the lines represent a 10-qubit circuit with 50-CX gates and 50 1-qubit gates, simulated on 10 different levels of noise.
The blue line `Original Farthest' represents a circuit in which all the 1-qubit gate angles are midway between Cliffords i.e., they have gate rotations: $\theta_{i} = n_{i}*(\pi/2) + \pi/4$ --- hence these angles are farthest away from Cliffords. 
The red line `Original Near' has the same circuit structure as above, but all the 1-qubit gates have only a small deviation from Cliffords: $\theta_{i} = n_{i}*(\pi/2) + \epsilon $.
The green line `Closest Clifford' circuit has the same structure, but with all angles as the nearest Clifford to the above: $\theta_{i} = n_{i}*(\pi/2) $.
The yellow line `Random Clifford' circuit  has the same structure, but with all angles as random Cliffords: $\theta_{i} = rand(n)*(\pi/2) $.

The first point to note is that the `Random Clifford' only captures the Original circuit's fidelity trend very loosely --- this alone is insufficient for our goals --- this circuit is far from the Original, with insignificant state overlap.
On the other hand, the `Closest Clifford' is excellent at capturing the fidelity trend, closely mimicking the two Original circuits.
Unsurprisingly, it almost overlaps the red line, since the angle difference (and hence the output state difference) between the two is marginal, but most importantly it also closely captures the trend of the blue line, which is the state that is farthest away from it.
This is clearly indicative that even though the states of the two circuits are not the same, the impact of structural noise is significant enough to create  similar fidelity trends as long as there is some non-trivial  state overlap. 

\begin{figure}[t]
\centering

\includegraphics[width=0.95\columnwidth,trim={0cm 0cm 0cm 0cm},clip]{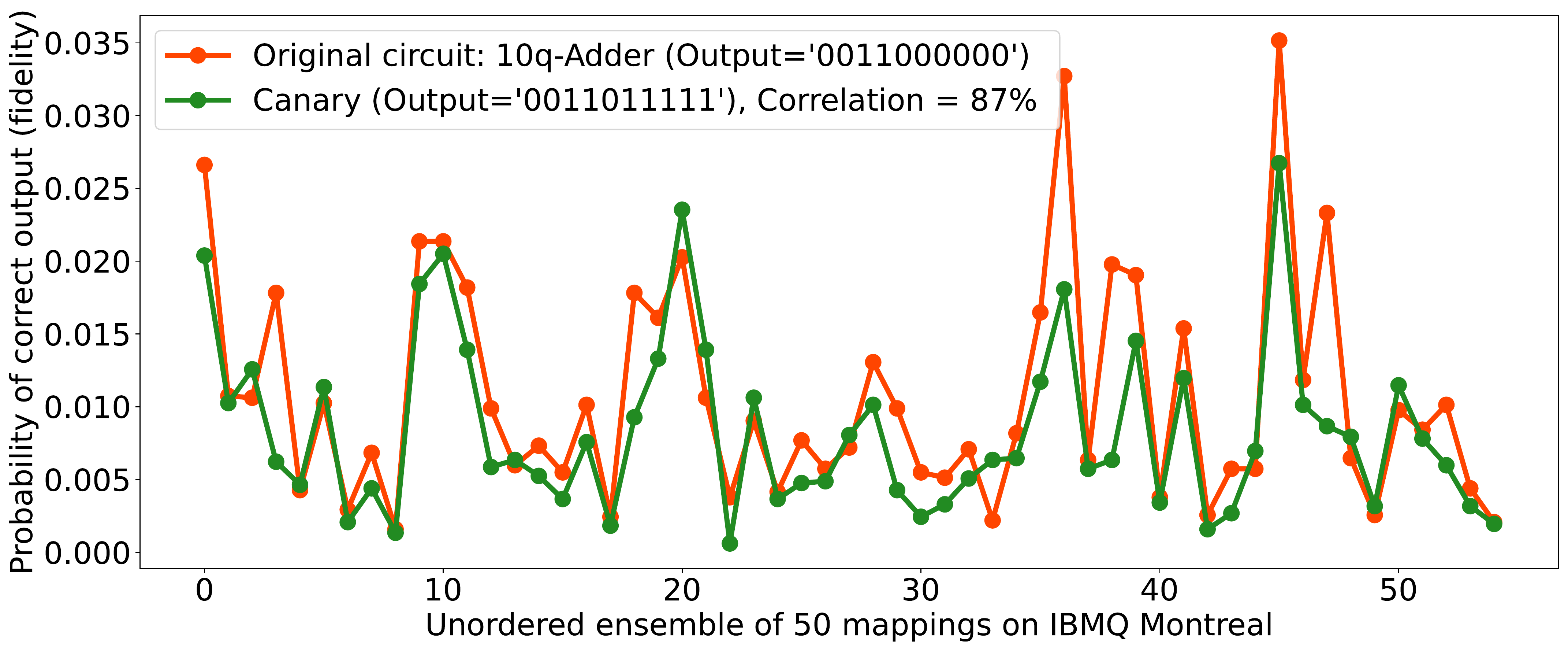}
\caption{High correlation in ensemble-wide fidelity between a 10q adder (orange) and its canary (green), across an ensemble of 50  mappings on IBM Quantum Montreal. %Though the output produced by the circuits are different (and at low fidelity), the two trends are fairly similar across the ensemble. 
% The correlation between the two is 87\%, indicating that the canary is highly accurate in predicting a good ensemble order for the original circuit. %TODO CAMERA
}
\label{fig:quancorde_cliff_1}
\end{figure}

\subsection{Quantifying canary ordering capabilities}
\label{e:quant_can}

Now, we show the capability of the canary circuits to capture fidelity trends of real applications on real devices.
Fig.\ref{fig:quancorde_cliff_1} shows the high correlation in ensemble-wide fidelity between a 10-qubit adder circuit (orange) and its corresponding canary circuit (green) generated by the methodology described earlier.
The comparison is performed across an ensemble of 50 different mappings on IBMQ Montreal.
Note that, while this figure assumes that the output of the original adder circuit (and therefore its fidelity) is known, this would not be the case for real-world quantum applications.
%This original output is only used here to show the `goodness' of the canary circuit.

First, observe that the outputs produced by the two circuits are different (as shown in the legend) - this is expected from prior discussion.
%This is expected because changing the 1-qubit gates when generating the canary changes the circuit state produced by the circuit in comparison to the original adder.
Second, note that the fidelity of both the original and the canary circuits are very low, under 4\% - these are hard to execute circuits on today's quantum devices.
Third, within this ensemble, it is unclear which mapping will produce the highest fidelity. It is not the `optimal' mapping chosen by the Qiskit transpiler, which is the first instance shown in the figure, and is clearly lower in fidelity compared to a couple of other instances.
Though the output produced by the circuits are different and at low fidelity, maintaining the same circuit structure (as seen in Fig.\ref{fig:quancorde_proposal_clifford}) enables the circuits to experience similar noise effects across the ensemble.
The two trends are fairly similar across the ensemble, even if there are some instances of non-negligible deviation. 
The fidelity correlation between the two circuits is 87\%, indicating that the canary is highly accurate in predicting a good ensemble order for this relatively low-fidelity original circuit.

\begin{figure}[t]
\centering
\includegraphics[width=0.95\columnwidth,trim={0cm 0cm 0cm 0cm},clip]{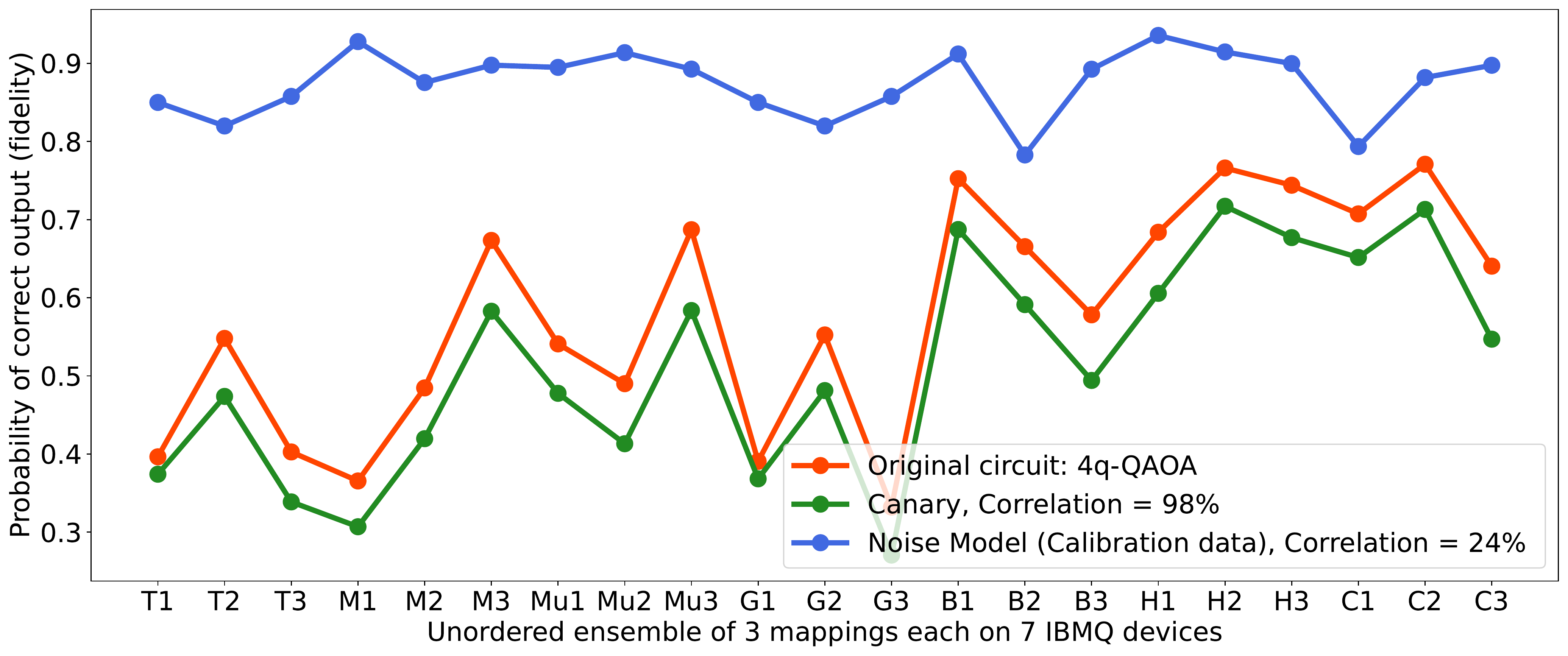}
\caption{ A 4-qubit QAOA circuit's fidelity across an ensemble of 3 mappings each on 7 IBM Quantum machines is shown (orange). Fidelity of the corresponding canary is also shown (green) and achieves near-perfect correlation with the original. Also shown is the  poorly  predicted  fidelity by a simulated noise model (blue) that uses the day's machine calibration data. 
%Clearly, the noise model's fidelity trend is poorly correlated (only 24\%) with the original, even for a simple circuit. %TODO CAMERA
%This clearly motivates using canaries to order the ensemble.
}
\label{fig:quancorde_cliff_2}
\end{figure}

Although it is evident from Fig.\ref{fig:quancorde_cliff_1} that the canary circuit can produce a good ensemble ordering for the target circuit, it would be fair to ask if an equally good ordering can be produced by simpler techniques, such as simply using the static noise information about the circuit.
After all, the mapped qubits are known and the corresponding coherence times and gate errors are available from the device's most recent calibration (which is roughly performed once a day for IBM Quantum machines).
Perhaps this noise data can be used to construct some fidelity heuristic?
%similar to the discussion in Section \ref{p:ens} (which simulated a simple 2-qubit circuit).
We argue that this is insufficient for fairly complex circuits and/or real device execution, for two reasons.
First, the noise data obtained from the devices are stale - the devices are characterized only at coarse granularities of time (usually at the time of calibration) because per-qubit characterization is very time-consuming, and the noise characteristics of the qubits tend to drift over time~\cite{Proctor_2020}.
Second, even if noise data were accurate, it is non-trivial to actually obtain the impact of these noise characteristics on the fidelity of a circuit of any reasonable complexity. 
This would require understanding every stage of the circuit's execution, which is as complex as the circuit execution itself (and thus classically inefficient). 

To quantify this argument, we perform an experiment with a fairly simple 4-qubit QAOA circuit in Fig.\ref{fig:quancorde_cliff_2}.
The fidelity of the circuit across an ensemble of 3 mappings, each on 7 IBM Quantum machines, is shown in orange. 
The fidelity of the corresponding canary circuit across the ensemble is shown in green and is seen to achieve near-perfect correlation with the original circuit.
This circuit is of low width and depth, and this is a fairly easy task for the canary. 
In addition, the fidelity predicted by the Qiskit simulation noise model is shown in blue.
This noise model uses the day's calibration data from the machines' qubits and gates. 
Clearly, the fidelity trend from the noise model is poorly correlated (only 24\%) with the original, even for a simple circuit.
The peaks and valleys of the original circuit's fidelity across the ensemble are not captured and, thus, this is poor for ensemble ordering.
%This is in line with the prior discussion.
%As discussed earlier, this is potentially due to drift in noise as well as limitations to perfectly modeling the real noise impact on fidelity. 
Further, we argue that this noise model based simulation is a loose upper bound on the accuracy that any noise based fidelity heuristic can produce. 
A more simple heuristic might multiply different error rates and coherence times, whereas the noise model is actually simulating the circuit with the noise data.
Note that this noise model simulation is itself not a scalable approach and is only illustrative for simple circuit examples.
Thus, it is clear that noise data based heuristic predictions of fidelity are inaccurate even for simple circuits, clearly motivating the canary approach to order the ensemble.

\begin{figure}[t]
\centering
%\fbox{
\includegraphics[width=0.98\columnwidth,trim={0cm 0.3cm 0cm 0cm},clip]{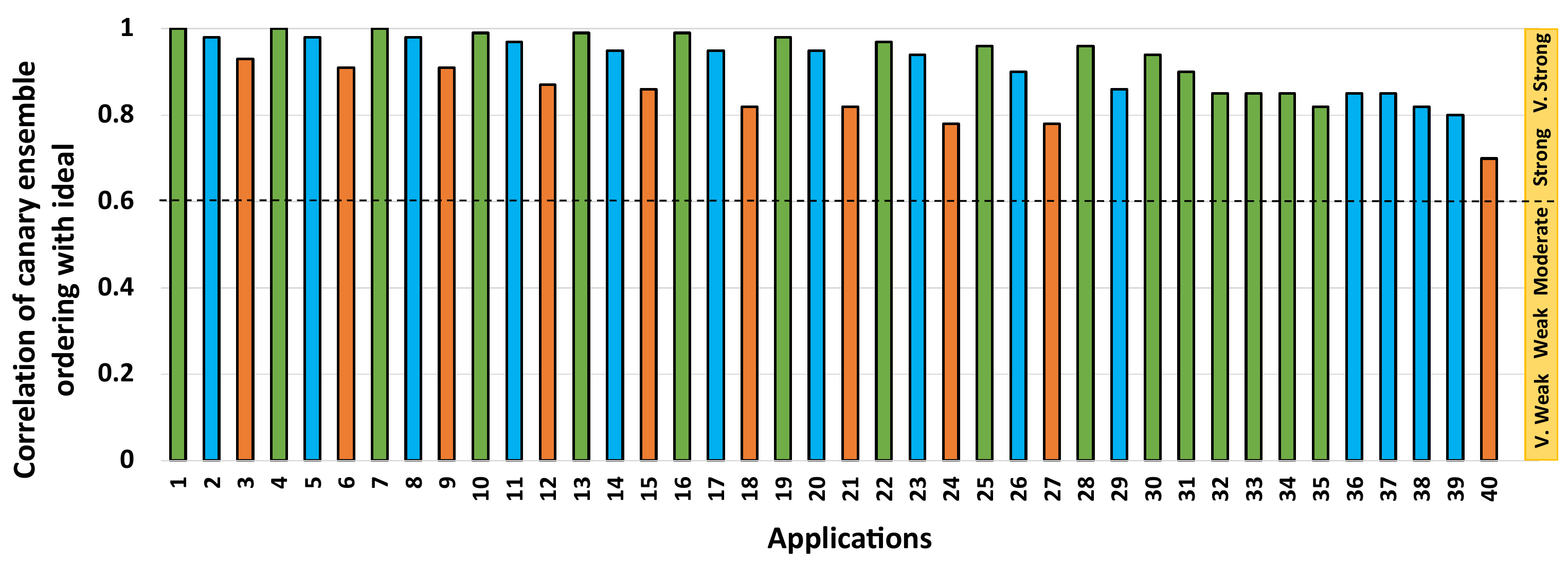}%}
\caption{Spearman correlation (varies from -1 to 1) between the canary-produced ensemble ordering and the ideal ordering for the target circuit, shown for 40 circuits (from real machine execution). The circuits are ordered by increasing size from 4-14 qubits.  Green (fidelity $> 10 \%$) / Blue  ( $2-10\%$) / Orange  ( $< 2\%$). Circuits always show (very) strong correlation.
%Even the lowest correlation observed (around 0.5) is still effective in producing reasonable ensemble ordering. %TODO CAMERA
%that helps in identifying / shortlisting correct outputs for the target circuit.
}
\label{fig:quancorde_cliff_3}
\end{figure}

Next, Fig.\ref{fig:quancorde_cliff_3} quantifies the high-accuracy ensemble orderings that our canary circuits are able to produce for a variety of applications. 
The figure shows the Spearman correlation between the canary-produced ensemble order and the ideal ensemble order for the target circuit (which would be unknown for complex non-simulable circuits), for 40 circuits.
The circuits are ordered by increasing size from 4-14 qubits.  Green (fidelity $> 10 \%$) / Blue  ( $1-10\%$) / Orange  ( $< 1\%$).
Spearman correlation measures the strength and direction of monotonic association between two variables and varies from -1 to 1.
Positive values, especially those closer to one, indicate a very strong positive correlation.
Correlations for all circuits are in the strong to very strong range, clearly indicating the effectiveness of Clifford canaries on real machines, even for circuits which are nearly at the brink of random output distributions. 
%Moving forward we focus only on the medimum and larger circuits, since the correct outcomes for the smaller circuits are fairly easy to decipher without Quancorde.

\iffalse
Three points to note.
First, even the lowest correlation observed (around 0.5) is still effective in producing reasonable ensemble ordering that helps in identifying and shortlisting correct outputs for the target circuit.
Second, one of the primary contributors to low correlation is the fact that the fidelity for large circuits is extremely low, and these are therefore extremely sensitive to dynamic variations (which could affect the original and canary circuit executions differently).
As machines improve, we will expect the correlations of the larger circuits chosen here to improve as well.
Third, correlation will also improve as the diversity and size of ensembles increase, since there will be greater `uniqueness' in the better ensemble orders.
\fi

\begin{figure}[t]
\centering
\includegraphics[width=0.95\columnwidth,trim={0cm 0cm 0cm 0cm},clip]{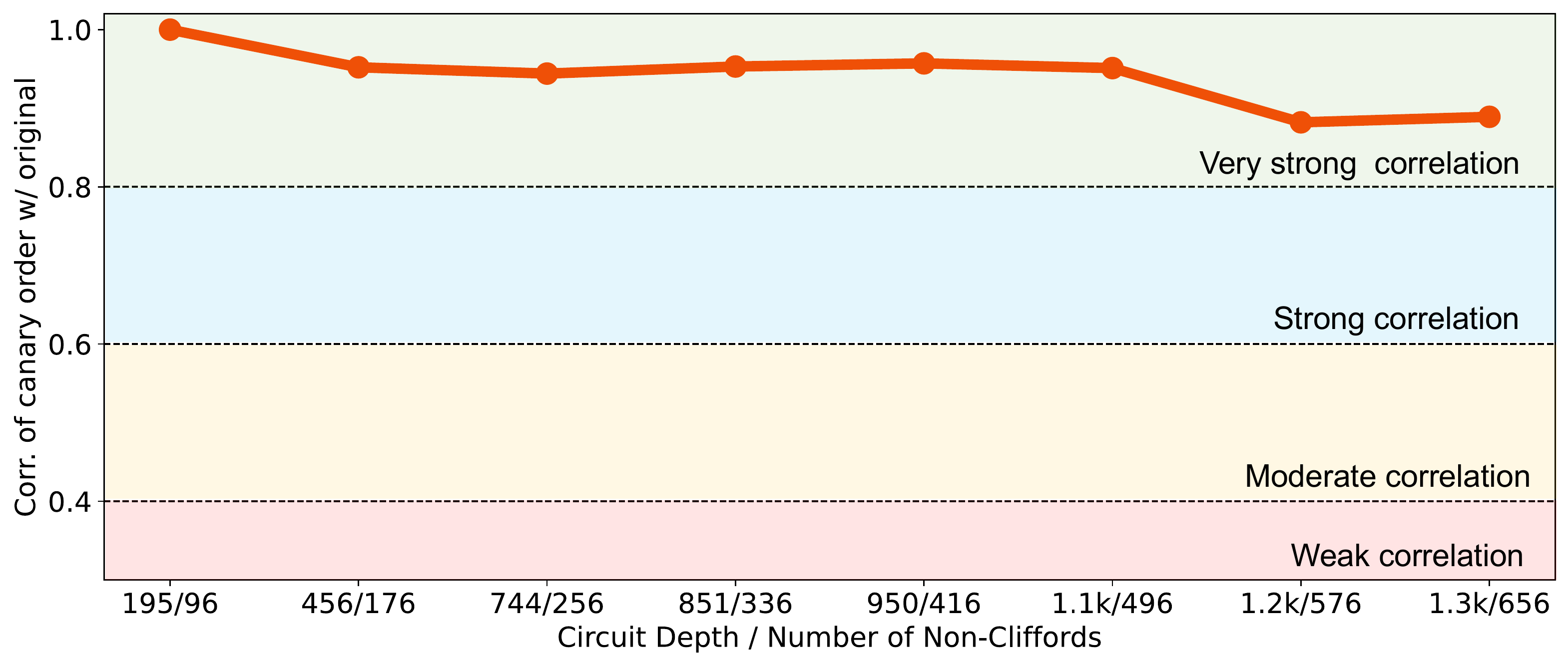}
\caption{Spearman correlation (varies from -1 to 1) between the canary-produced ensemble ordering and the ideal ordering for the target circuit, shown for 8 deep 12q circuits with many gates (in simulation). %Very strong correlation is observed even for circuits of 1,300 CX gate depth / 650 Non-Clifford gates.
}
\label{fig:quancorde_cliff_corr_sim}
\end{figure}

\subsection{Scaling to more complex circuits}
We wish to show that canary circuits can accurately follow the fidelity trends of original circuits even when the number of non-Clifford gates grows considerably.
Unfortunately, on today's quantum devices, circuit which are deeper than 100-200 CX gates and wider than 10-20 qubits begin to produce purely random distributions which no post-processing mitigation technique can improve.
Thus, to evaluate the capabilities of the Clifford canaries on more complex circuits, we perform experiments in noisy simulation on very deep, reasonably wide, circuits. 
In Fig.\ref{fig:quancorde_cliff_corr_sim}, we show 12-qubit circuits with circuit depth ranging from 195-1,300 CX gates and with the total number of non-Clifford gates ranging from 96-656.
It can be observed that the correlation between the Clifford canaries and the original circuits is always in the very strongly correlated range.
This is a clear indicator of the scalability of this approach to more complex circuits, as and when they can be run on quantum devices with non-negligible fidelity. 
Note that the correlations seen here are generally higher than those in Fig.\ref{fig:quancorde_cliff_3} - this is because real machine noise is more erratic than noisy simulation.
Nevertheless, the utility of the Clifford canaries is evident.

\section{Methodology}
\label{sec:method}

% \usepackage{graphicx}
\begin{table}[]
\resizebox{\columnwidth}{!}{%
\begin{tabular}{|l|l|l|l|l|l|}
\hline
\textbf{App} & \textbf{Q} & \textbf{Output} & \textbf{Depth} & \textbf{MaxFid} & \textbf{AvgFid} \\ \hline
QAOA6\_1 & 6  & 101011+011000  & 41  & 29\%  & 11.0\% \\ \hline
ADD6\_1  & 6  & 110000         & 29  & 32\%  & 10.0\% \\ \hline
ADD6\_2  & 6  & 111100         & 29  & 25\%  & 9.0\%  \\ \hline
QFT6     & 6  & Equal superposn.         & 36  & 42\%  & 7.0\%  \\ \hline
QAOA6\_2 & 6  & 011001+001000  & 93  & 10\%  & 5.0\%  \\ \hline
ADD8\_1  & 8  & 00111100       & 41  & 13\%  & 3.7\%  \\ \hline
QAOA6\_3 & 6  & 001001+001000  & 243 & 5\%   & 3.7\%  \\ \hline
QAOA6\_4 & 6  & 011001+001000  & 283 & 6\%   & 3.7\%  \\ \hline
ADD8\_2  & 8  & 11111100       & 41  & 11\%  & 2.6\%  \\ \hline
ADD10\_1 & 10 & 0011000000     & 53  & 3.5\% & 1.0\%  \\ \hline
QFT8     & 8  & Equal superposn.       & 52  & 1.7\% & 0.9\%  \\ \hline
ADD10\_2 & 10 & 1111110000     & 53  & 2\%   & 0.4\%  \\ \hline
%QFT10    & 10 & 0000000000     & 68  & 0.8\% & 0.6\%  \\ \hline
ADD12\_1 & 12 & 000011001100   & 65  & 0.7\% & 0.5\%  \\ \hline
ADD14\_1 & 14 & 00000011110000 & 77  & 0.5\% & 0.4\%  \\ \hline
ADD12\_2 & 12 & 111111111100   & 65  & 0.5\% & 0.35\% \\ \hline
\end{tabular}%
}
\caption{Evaluated Adder, QAOA, and QFT benchmarks. Circuit outputs, pre-mapping CX depth, maximum fidelity, and average fidelity across all ensembles are shown.}
\label{tab:method}
\end{table}

\subsection{Applications}
\label{5-m-circ}
%TimeStitch aims to apply quantum theory and NISQ machine insight to optimize the scheduling of quantum circuits. 
Our benchmarks are representative of real-world usecases, and are described below and detailed in Table \ref{tab:method}.
All applications have average baseline fidelity at 10\% or lower - i.e., we are primarily interested in applications for which correct outcomes are relatively hard to identify. 
Note that maximum fidelity can be greater, but the machine / mapping producing the best-case fidelity is unknown and often non-intuitive.

%TODO CAMERA
%Due to limitaions on circuit width because of machine size and depth because of coherence times on available IBM QCs, benchmarks that included 6 qubits or fewer and of shorter duration were included for evaluation. 
%Brief descriptions of the benchmarks used in our study are as follows, and their characteristics are listed in Table \ref{tab:method}.

\emph{Ripple Carry Adder:}
%TODO CAMERA
Adders are a critical building block for quantum logic. 
%As they generate known outputs, it was chosen to assist with the evaluation of our results. 
We implement linear-depth ripple-carry adder quantum circuits for 2-6 bits that utilize 6-14 qubits respectively~\cite{cuccaro2004new}. 
Further, each of these are evaluated for multiple input/output bitstrings.

\emph{Quantum Fourier Transform:} 
QFT is a circuit used as a building block for applications such as Shor's factoring and phase estimation.
It converts a quantum state from the computational basis to the Fourier basis through the introduction of phase. 
%We evaluated a concatenated circuit of QFT + QFT$^{-1}$ that produces the ground state as output.
QFT was constructed for 6-8 qubits~\cite{mike_ike_2020}.

\emph{Quantum Approximate Optimization Algorithm:} 
QAOA \cite{farhi2014quantum} is a variational quantum-classical algorithm to solve combinatorial optimization problems.
%TODO CAMERA
QAOA is implemented atop a parameterized circuit called an ansatz.
%and we use one instance of a hardware efficient QAOA ansatz as the benchmark.
%and its solution is simple to predict when solving MAXCUT on a ``ring of disagrees'' graph structure. 
We use QAOA ansatz constructed for 6 qubits for 4 different input graphs (which result in different circuit depths).

\subsection{Infrastructure / Overheads}
\label{Infra}
Quancorde is implemented to interface with Qiskit~\cite{Qiskit} and is evaluated on IBMQ~\cite{IBMQS} devices. Canary circuits are efficiently constructed on classical compute.

\textbf{Intra-device evaluations:} Our intra-device evaluations target the 27-qubit IBM Quantum Montreal and we run 50 different circuit instances across the device.
One instance is that which is mapped and routed to the device by the Qiskit transpiler when run with highest optimizations enabled. 
%This often does \emph{not} produce the highest fidelity (again implying the limitations of noise data based heuristics), though it is usually reasonable.
The other 49 are random mappings. % which can be a different subset of qubits within the device or a different mapping on the same set of qubits.

\textbf{Inter-device evaluations:} Our inter-device evaluations target 7 IBM Quantum devices: Montreal (27q), Toronto (27q), Guadalupe (16q), Hanoi (27q), Cairo (27q), Mumbai (27q) and Brooklyn (65q).
Only one circuit is run per device, mapped and routed to the device by the Qiskit transpiler when run with highest optimizations enabled.

\textbf{Clifford simulation:} Clifford circuits can be run on any stabilizer simulation framework on any classical computer - we use the IBM stabilizer framework which is currently provisioned to  run up to 5000 qubits.
%As discussed earlier, these Clifford circuits of any size can be classically simulated in polynomial time (nearly linear). This is the Gottesman-Knill's theorem~\cite{gottesman1998heisenberg}.
Our target circuits run within seconds (including I/O etc).
Canary construction time is negligible (linear in the number of non-Clifford gates).

\textbf{Compilation time / shots:} The compilation time for the additional circuits is negligible since near-term quantum circuits can be compiled rapidly~\cite{QCloud}. All circuits are compiled in under a minute.
Circuit executions are run for 8192 (maximum number) execution shots.

\iffalse
\textbf{Ensemble overhead:} Unless specified, we used a fixed ensemble size for inter-machine / intra-machine experiments. 
More ensembles can be beneficial, but the size of the ensemble is not directly related to the size of the circuit and is more tied to the baseline fidelity.
For instance, as machine quality improves, deeper / wider circuits can sufficiently be boosted by Quancorde even with small ensembles. 
%Additional resource overheads are proportional to the size of the ensemble, but this is a fair price to pay for the critical benefit of quantum fidelity improvement. 
%All additional quantum and classical executions are performed in parallel to the baseline, so there are no time overheads.
Ensemble size analysis is performed in Section \ref{e:size}.
\fi

\textbf{Cloud execution:} The entire original+canary circuit ensemble (for each device) is run as a single batch. Quantum devices in the cloud are provisioned to run a batch of quantum circuits back to back with no wait time (as a single monolithic entity in the system queue). 
Thus, the canary and original are run back to back separated by roughly a millisecond or less.  
The effects of machine drift are minimal in this time frame. 
%Random transient errors can of course occur, but they are very rare and they would affect any execution / mitigation technique, not just Quancorde. Avoiding transients is a different area of research by itself.

\textbf{Correlation post-processing:} 
The correlation processing step involves calculating the correlation of the ensemble ordering for each output string in the baseline probability distribution with the canary ordering.
At first glance, it may seem that $2^n$ output strings are possible, but this is \emph{not} the case and the cost is not exponential. 
There are two upper bounds that can be established on the number of strings which have to be analyzed.
The trivial upper-bound is $shots*ensemble\_size$ i.e., the number of unique strings produced cannot be more than number of shots run over all the devices - this number would be reached if each device produced a purely random distribution, and each machines distribution was mutually exclusive. The number of shots is not exponential (if it were, quantum computing would be infeasible) and can be derived from the target decipherable fidelity on the quantum devices.
There is an even lower upper-bound that can be derived directly from a minimum baseline fidelity target. If we establish that the minimum application fidelity we are interested in boosting is, say, $f=0.1\%$ (since in this region, the output distribution starts becoming purely random), then we would at most have to analyze only the top $1/f$ (= 1000 for $f=0.1\%$) unique strings in terms of their probability of occurrence. This is because there cannot be more than 999 strings which have probability of occurrence greater than or equal to than the correct outcome, because if it were so, that would mean that the fidelity would have to be less than 0.1\%. This can be trivially reasoned mathematically.
\iffalse
To show this, assume the correct outcome had probability p, let's say another 1000 strings had probability p+e for a tiny e. Let's say all the other strings had negligible probability. Then fidelity(correct outcome) = p/(sum of probability of all strings) < p/(probability of top 1001 strings) = p/(p+ 1000*(p+e)) < p /(p+1000*p) = 1/1001, which is less than 0.1\% . This violates our target fidelity.
\fi
Thus, the correct outcome will have to be among the top 1000 strings. Or more generally, among the top $1/f$ strings for a target lower bound baseline fidelity $f$. This reasoning can similarly be extended to multi bit-string scenarios as well, but again, clearly it is non exponential.
With this fidelity target of 0.1\%, we evaluate the actual number of strings in Section \ref{rank-string} - we find the numbers to be well within the bounds.

\iffalse
It should be noted that $2^n$ correlation comparisons of all possible output string orderings, with the canary ordering, is \emph{not} required. 
At most only N (= $1/min\_baseline\_fidelity$) number of comparisons are required. 
This is because, if we define some minimum baseline fidelity up to which Quancorde is useful, then the correct string cannot be further than N strings away away from the top, i.e., its rank has to be at most N. Thus, comparing those top N strings is sufficient to find the right one.
For applications with greater than 0.1\% fidelity (our informal lower bound), N=1000 is sufficient - this is conservative since distributions are considerably better than random. 
Note that this number is practically a constant and does not increase with the number of qubits or circuit depth - it is only governed by the target lower bound on baseline fidelity .
Baseline fidelities that are lower cannot be mitigated by any data processing technique since they are effectively random distributions. 
\fi

\subsection{Evaluation Metrics}
\label{5-m-evalcompare}

%Since Quancorde is a fundamentally new approach, it requires relevant evaluation metrics.
We evaluate Quancorde benefits through 2 metrics:
%evaluate Quancorde on 3 metrics.
%The first is specific to the Quancorde approach.
%The second is a traditional metric that compares Quancorde against a highly optimized single device mapped baseline.

\circled{1}\ \textbf{\emph{Rank of correct outcome:}} This metric indicates the ranking of the correct outcome among the correlations of all possible outcomes with the canary-produced ensemble order. We refer to this as `Rank'. If Quancorde works perfectly, the rank of the correct outcome is 1. But for more challenging circuits, the rank is not always 1. Note that while a rank nearer to 1 is always more beneficial, it is not required for Quancorde's success - even lower ranks allow significant fidelity improvements over the baseline.
%preferred, but not required, for Quancorde's success.  We then perform a sensitivity analysis to analyze the worst rank among all our applications. Even our hardest-to-execute applications will have the correct outcome among the worst-rank number of selected output strings. 

\iffalse
\circled{2}\ \textbf{\emph{Relative correlation of correct outcome wrt. highest:}} This metric focuses on the correlation of outcome probability with the canary-produced ensemble order. It indicates the relative correlation value of the correct outcome compared to the correlation value of the most correlated outcome. We refer to this as `Relative Correlation' (RC). If Quancorde works perfectly, this value is 100\% (i.e., the correct outcome is the most correlated). Similar to the above, for more challenging circuits, Quancorde is not always equal to 100\%. We then perform a sensitivity analysis to analyze the worst of this metric. This tells us how far out from the highest outcome we would need to search the correlation distribution so that the correct outcome will definitely be among the searched output strings, even for our hardest-to-execute applications. 
\fi

\circled{2}\ \textbf{\emph{Fidelity:}} This metric is a standard evaluation metric that indicates the likelihood of finding the correct outcome of a quantum application. %Quancorde returns multiple outcomes per application (say `k') with the knowledge that the correct outcome will very likely be among these. Then the fidelity of that application is $(1/k)$ since the probability of choosing the correct outcome from the set is 1 among k.
We show the benefits of Quancorde against: a)  the maximum baseline fidelity obtained for a target application across all mappings and machines --- this is very favorable to the baseline since the best performing machine / mapping is technically not known (and, in our experiments, is often not the Qiskit optimized one) and, b) the average benefits across the entire ensemble.
Practical benefits can lie somewhere in between. %We can fairly compare this against the fidelity of a traditional baseline, which is basically the probability of choosing the correct outcome from among the total repeated executions (i.e., shots) of a circuit run on a single device / mapping. 

\section{Evaluation}
\label{sec:eval}
\iffalse

\begin{figure}[t]
\centering
%\fbox{
\includegraphics[width=0.98\columnwidth,trim={0.4cm 0cm 0.1cm 0cm},clip]{figures/Quancorde_bench_result_1_sort.pdf}
%}
\caption{\hl{Correlation of occurrence probability (across the ensemble) of each output string to the canary-produced ensemble order. This is shown for a 12-qubit adder circuit on an intra-machine ensemble with max baseline fidelity < 1\%. The correct adder outcome is most correlated with the ensemble order (Rank=1/4096), thus the correct adder outcome can be accurately identified.}}
\label{fig:eval_app_intra}
\end{figure}

\begin{figure}[t]
\centering
%\fbox{
\includegraphics[width=0.98\columnwidth,trim={0.3cm 0cm 0.1cm 0cm},clip]{figures/Quancorde_bench_result_2_sort.pdf}
%}
\caption{\hl{Correlation of occurrence probability (across the ensemble) of each output string to the canary-produced ensemble order. This is shown for a 10-qubit QFT+QFT$^{-1}$ circuit on an inter-machine ensemble with max baseline  fidelity < 1\%. The correct outcome has Rank=12/1024. Thus, returning 12 or more of the top correlated outcomes would contain the correct outcome.\todo{remove this - one is sufficient}}
\label{fig:eval_app_inter}
\end{figure}

\fi

\begin{figure}[t]
\centering
%\fbox{
\includegraphics[width=0.95\columnwidth,trim={0cm 0cm 0cm 0cm},clip]{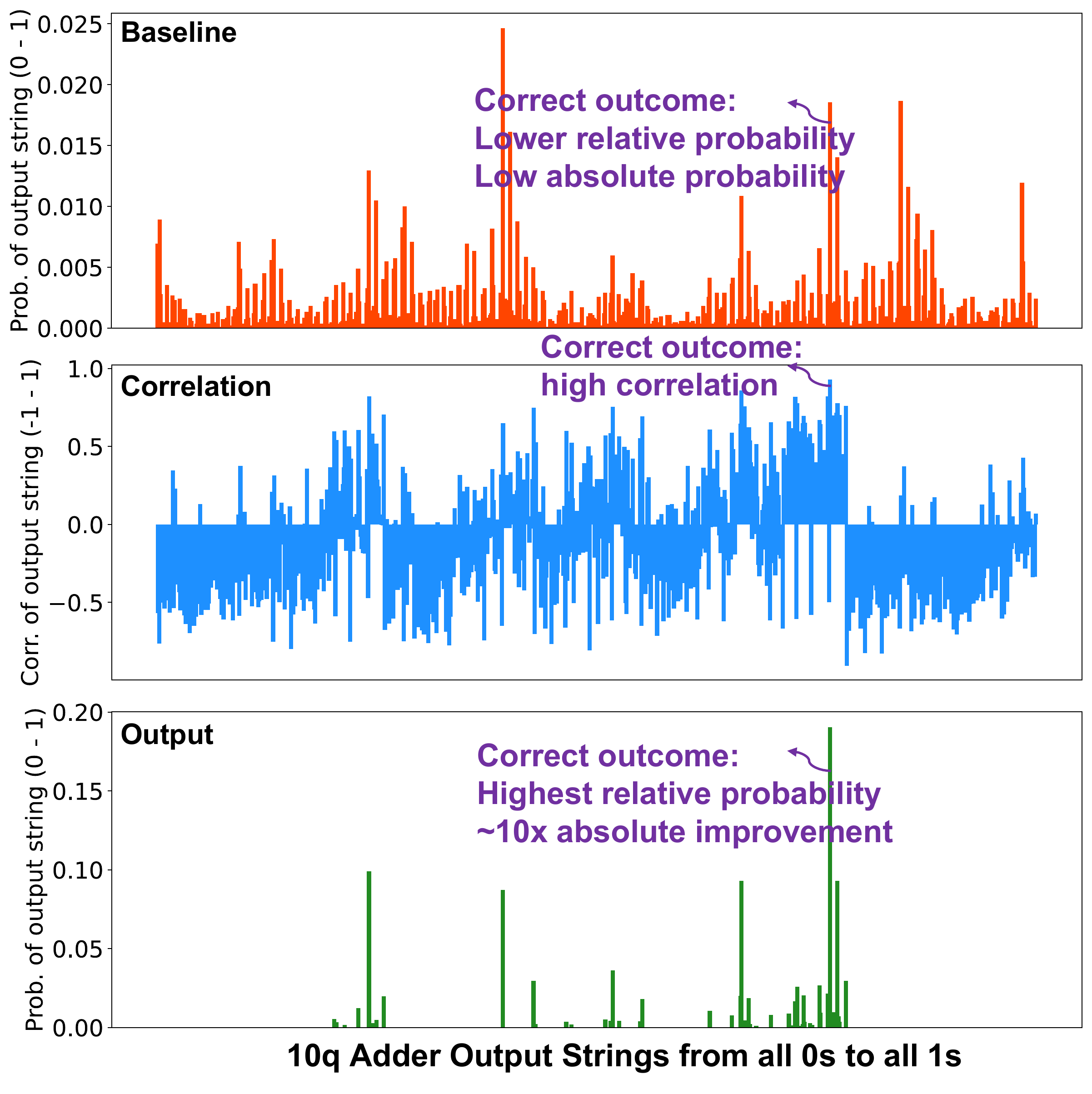}
%}
\caption{Bitstring distribution produced by the baseline for a 10-q adder is shown in red. The probability of the correct outcome is low in magnitude (2\%) and also lower than other strings.
The Correlation distribution produced by Quancorde (blue),  with an intra-machine ensemble, shows the correct outcome to have the highest correlation with the canary  ensemble.
When this is used to weight the original distribution, the  Output distribution produced (green) has the correct outcome at 20\% probability and is a clear winner. 
}
\label{fig:eval_app_rev}
\end{figure}

\subsection{Single-application 10-qubit adder analysis}
%First, we re-emphasize that Quancorde does not require the correct outcome to have the highest baseline probability among the all produced bitstrings. This is the fundamental per-machine limitation that Quancorde is able to overcome. 

%Figures \ref{fig:eval_app_intra} and \ref{fig:eval_app_inter} show examples of the benefits produced by Quancorde.
Fig.\ref{fig:eval_app_rev} is a real machine / application example of the benefits produced by Quancorde, with an intra-machine ensemble. 
It shows the analysis for a 10-qubit adder which has a baseline machine fidelity of under 1\% - the correct answer is a single bitstring `1111110000'. 
In all 3 graphs, the x-axis shows all the different 10-bit strings.
The top figure shows the output probability distribution for the baseline.
The correct outcome is indicated.
Note that the probability of its occurrence (i.e., fidelity) is very low, only 1.7\%.
Further, it is not the output string with the highest probability.
Clearly, the correct outcome cannot be identified.
The middle figure shows Quancorde's correlation analysis - the correlations vary from -1 to 1, with near 1 indicating high positive correlation.
The correlation of the correct outcome is indicated and it has the highest correlation of nearly 0.9.
%Note there are some other strings with high correlation as well - this is okay.
Next, the bottom figure shows the weighted output distribution by incorporating the correlation into the baseline distribution.
Only a few strings remain, and the correct outcome's probability is boosted by 10x, to 20\%. 
Further, it is a clear winner among all the strings.

\subsection{Outcome Ranks and Correlation Strings}
\label{rank-string}

\iffalse
\begin{figure}[h]
\centering
%\fbox{
\includegraphics[width=0.98\columnwidth,trim={0cm 0cm 0cm 0cm},clip]{figures/QC_Full_Eval_1.pdf}
%}
\caption{Rank and RC produced by Quancorde for 16 applications with an intra-machine ensemble. Most small and medium sized circuits have perfect Rank and RC. The hard-to-execute circuits produce substantially good Ranks and high RC. 
%The Rank and RC mean and worst-case are 2.46/98.6\% and 30/95\%, respectively. %TODO CAMERA
}
\label{fig:eval_full_intra}
\end{figure}
\fi

\iffalse
\begin{figure}[h]
\centering
%\fbox{
\includegraphics[width=0.98\columnwidth,trim={0cm 0cm 0cm 0cm},clip]{figures/QC_Full_Eval_2pdf.pdf}
%}
\caption{Rank and RC produced by Quancorde for 16 applications with an inter-machine ensemble. Again, small and medium sized circuits have mostly perfect Rank and RC. The hard-to-execute circuits' results are also promising, and slightly better than the intra-machine scenario \todo{Perhaps remove this and keep only the other one - can argue that we focus on intra because of less resource cost}. 
%The Rank and RC mean and worst-case are 2.16/98.8\% and 15/95.5\%, respectively. %TODO CAMERA
}
\label{fig:eval_full_inter}
\end{figure}
\fi

\begin{figure}[t]
\centering
%\fbox{
\includegraphics[width=0.95\columnwidth,trim={0cm 0cm 0cm 0cm},clip]{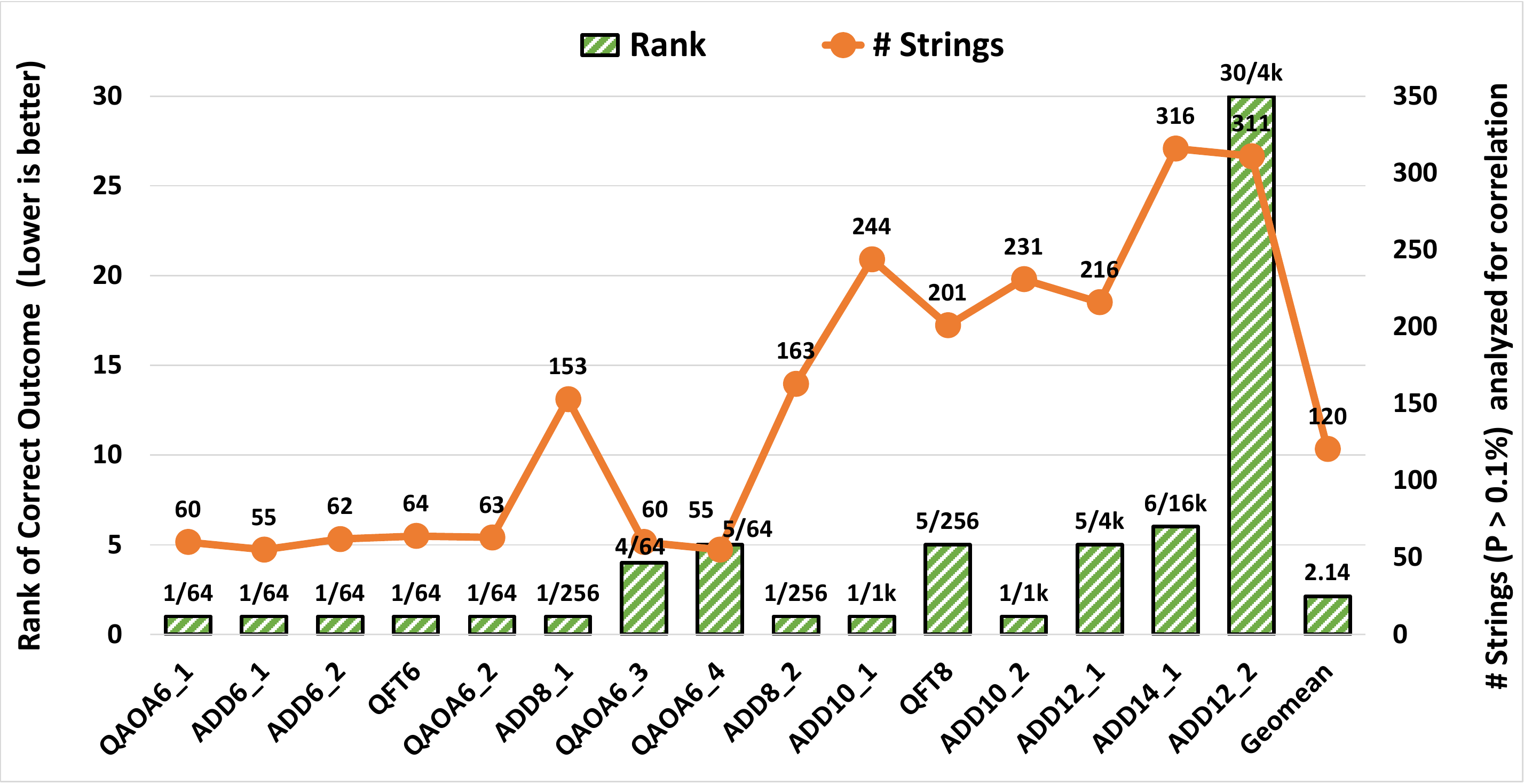}
%}
\caption{Left axis / bars: Rank produced by Quancorde for 15 applications. Right axis / line: Number of output strings (with occurrence probability greater than 0.1\%) analyzed for correlation.
%Most small and medium sized circuits have perfect Rank. The hard-to-execute circuits produce substantially good Ranks and high RC. 
%The Rank and RC mean and worst-case are 2.46/98.6\% and 30/95\%, respectively. %TODO CAMERA
}
\label{fig:eval_full_intra}
\end{figure}

Figure \ref{fig:eval_full_intra} shows two Quancorde results for 15 applications. The results are averaged across 3 trials of intra-machine and inter-machine ensembles. 

First, we look at the left axis / green columns, which shows the rank of the correct outcome in the correlation distribution.
As discussed in Section \ref{5-m-evalcompare}, a rank closer to 1 can improve fidelity further, since this would imply the correct outcome having a higher correlation compared to other outputs, thus boosting it further in the final output distribution.
But lower ranks can still produce substantial benefits. 
It is evident that many of the relatively smaller circuits (6 qubits) have perfect Rank.
The larger more complicated circuits (8 qubits and greater) have Ranks in the range of 1-30.
Even the worst-case Rank 30 occurs for a distribution which has 4096 potential outputs, clearly showcasing the potential for harnessing correlation.
%For example, in Fig.\ref{fig:eval_app_inter}, the 12-qubit adder ADD12\_1 impressively has a Rank of 1 out of 4k outcomes, and the 14-qubit adder ADD14\_1 has a Rank of 10 out of 16k outcomes. 
%The 12-qubit adder ADD12\_2 in Fig.\ref{fig:eval_app_intra} has the worst Rank in our evaluations -  30 out of 4k outcomes, which is still impressive.
Mean ranks across the applications show that the correct outcome usually occurs in the top 2-3 outputs in the correlation distribution.
% whereas the worst-case is 30. 
%Looking at RC, we see a mean of 98+\% across the two analyses, meaning that the correct outcome has a correlation coefficient within 2\% lower than the maximum.
%The worse-case bound across our applications is 5\%.
%It is important to note that the worst case instances occur for applications which have fidelity substantially lower than 1\%.
%Thus, even returning a reasonable set of outcomes can be very beneficial for further evaluations as discussed in Section \ref{sec:practical}. 
Also, we noted that Quancorde performs marginally better on inter-machine diversity compared to intra-machine diversity.
The difference is small enough to suggest that any form of diversity is useful.% to the Quancorde approach.

Next, we look at the right axis / orange line, which shows the number of strings analyzed in correlation processing.
As discussed in Section \ref{sec:method}, with a target baseline fidelity lower bound of 0.1\%, the maximum strings analyzed would be 1000.
From the figure, we see that the actual number only goes up to 316.
The smaller circuits have nearly all their bitstrings with a fidelity greater than 0.1\% , whereas the large circuits, like the 14-qubit adder, have less than 2\% of their strings worthy of correlation processing.
This clearly shows that the cost of correlation processing is very small, making Quancorde a computationally scalable approach.

\subsection{Fidelity benefits}
\label{e:fid}

\iffalse
\begin{figure}[h]
\centering
%\fbox{
\includegraphics[width=0.98\columnwidth,trim={0cm 0cm 0cm 0cm},clip]{figures/Quancorde_fidelity_EDM.pdf}
%}
\caption{\hl{Fidelity improvements of Quancorde compared to the baseline. Results are averaged over inter- and intra-machine ensembles. Quancorde fidelity is calculated by selecting a threshold-based top subset of high correlation k outputs. The threshold is estimated via a sensitivity analysis, and results are also shown for a conservative 2x threshold. Quancorde achieves a mean 6.9x improvement in fidelity over the baseline, as well as a conservative improvement of 3x. Higher benefits (10-31x) are achieved for hard-to-execute applications which have baseline fidelity < 1\%. In comparison, EDM}~\cite{Tannu:2019b}\hl{ achieves a 1.1x average (max 1.6x) fidelity improvement.} \todo{only this in the result really needs to change}}
\label{fig:eval_full_fid}
\end{figure}
\fi

\begin{figure}[h]
\centering
%\fbox{
\includegraphics[width=0.98\columnwidth,trim={0cm 0cm 0cm 0cm},clip]{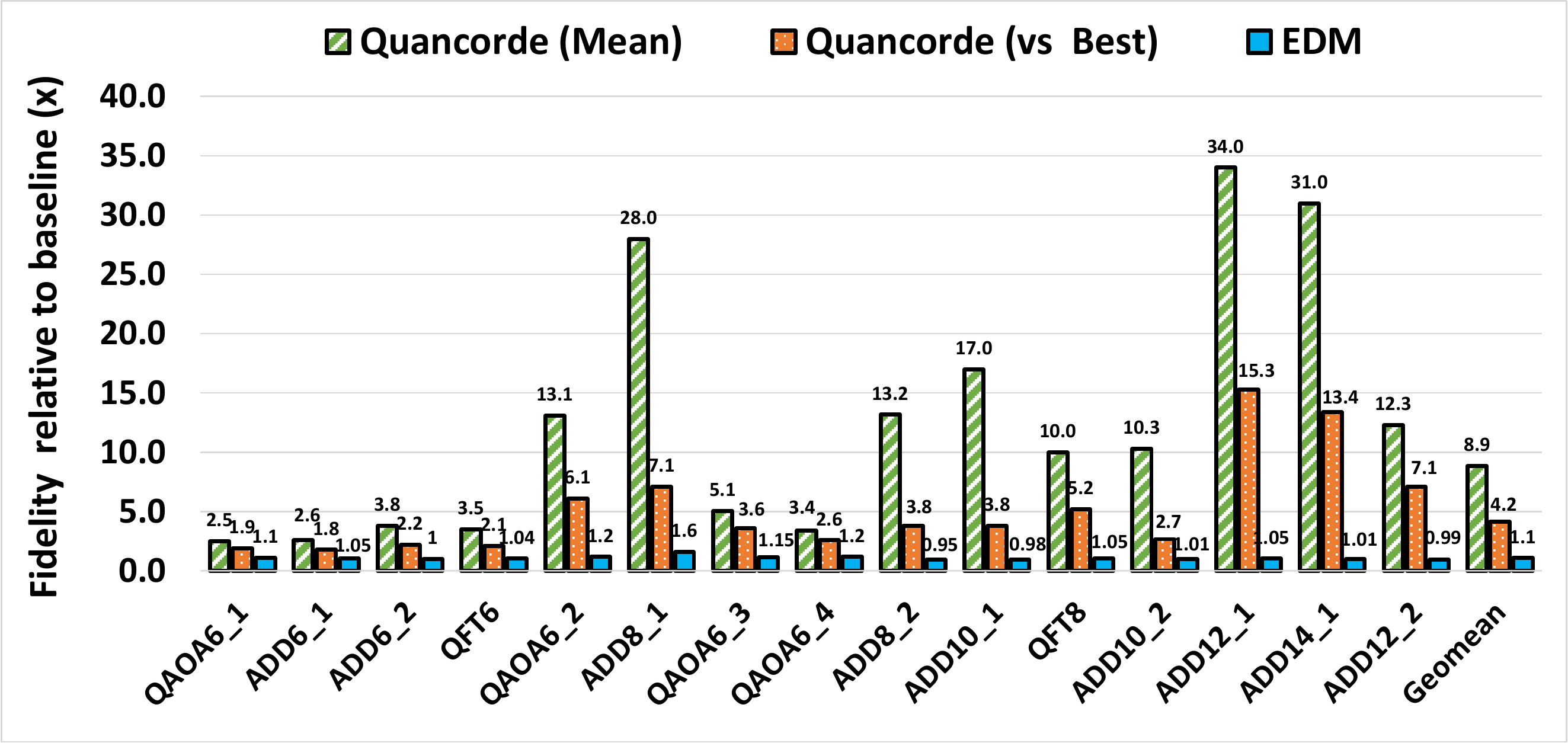}
%}
\caption{Fidelity improvements of Quancorde compared to the baseline. Results are averaged over inter- and intra-machine ensembles. Comparison to EDM is also shown.
%Improves are shown against the ensemble average as well as against the best from the ensemble. Quancorde achieves a mean 8.9x improvement and a 4.2x improvement over the best case. 
%Higher benefits are achieved for hard-to-execute applications which have baseline fidelity < 1\%. 
%In comparison, EDM~\cite{Tannu:2019b} achieves a 1.1x average (max 1.6x) fidelity improvement.
}
\label{fig:eval_full_fid}
\end{figure}

Next, we show the fidelity improvements of Quancorde compared to the baseline. Results are averaged over inter- and intra-machine ensembles.
Two sets of fidelity improvements are shown --- `Mean' and `vs Best' --- these were discussed in Section \ref{sec:method}.
Quancorde achieves considerable fidelity boosting, especially for the larger applications.
For smaller applications, the baseline mean fidelity is already around 10\% and as high as 42\%, so the boosting potential is limited. 
Nevertheless, Quancorde achieves very high fidelity after boosting.
For larger applications, Quancorde's benefits are very attractive. 
The improvements are as high as 15x / 34x (`Mean' and `vs Best' respectively) for the 12-qubit Adder, which had a correct outcome rank of under 5, out of 4k bitstrings.
On average, fidelity boosting of 8.9x / 4.2x is achieved, which is higher than typical error mitigation techniques, clearly highlighting Quancorde's novel capability of harness a diverse ensemble.
Note that considerable improvements are achieved even on uniform multi-output distributions like QFT.
Non-uniform distributions are discussed separately in Section \ref{e:multi}.

We also show comparison against EDM~\cite{Tannu:2019b}, another ensembling approach, albeit for a different goal of avoiding specific occurrences of correlated errors.
We use an ensemble size of 4 best mappings (which is the default setting). 
We observe fidelity improvements from EDM to be substantially lower than that obtained from Quancorde - 1.1x on average and a maximum of 1.6x.
EDM is seen to provide benefits primarily in scenarios of reasonably high baseline fidelity - the top 5-6 benchmarks in Table \ref{tab:method}.
To our knowledge, this is likely because EDM is suited to reducing the probability of occurrence of very specific competing incorrect bitstrings which have uncharacteristically high occurrence due to unique correlated errors. 
Such scenarios are unlikely near the threshold of device execution capability, at which point, the impact of all qubits, their gate errors and coherence times are all rather significant.
Using multiple mappings like EDM, while potentially cancelling out some correlated errors, can also have a negative impact on the correct bitstring, especially when the baseline fidelity is extremely low.
This is likely the reason for some applications suffering fidelity degradation with EDM.

A key takeaway from the above analysis is that Quancorde can improve application fidelity beyond the execution capability of any single device / mapping.
The relative benefits will improve as circuit fidelity decreases (alternatively, as circuit depth increases) until the point that the circuit output is still somewhat decipherable and not simply a random distribution.

\begin{figure}[t]
\centering
%\fbox{
\includegraphics[width=0.95\columnwidth,trim={0cm 0cm 0cm 0cm},clip]{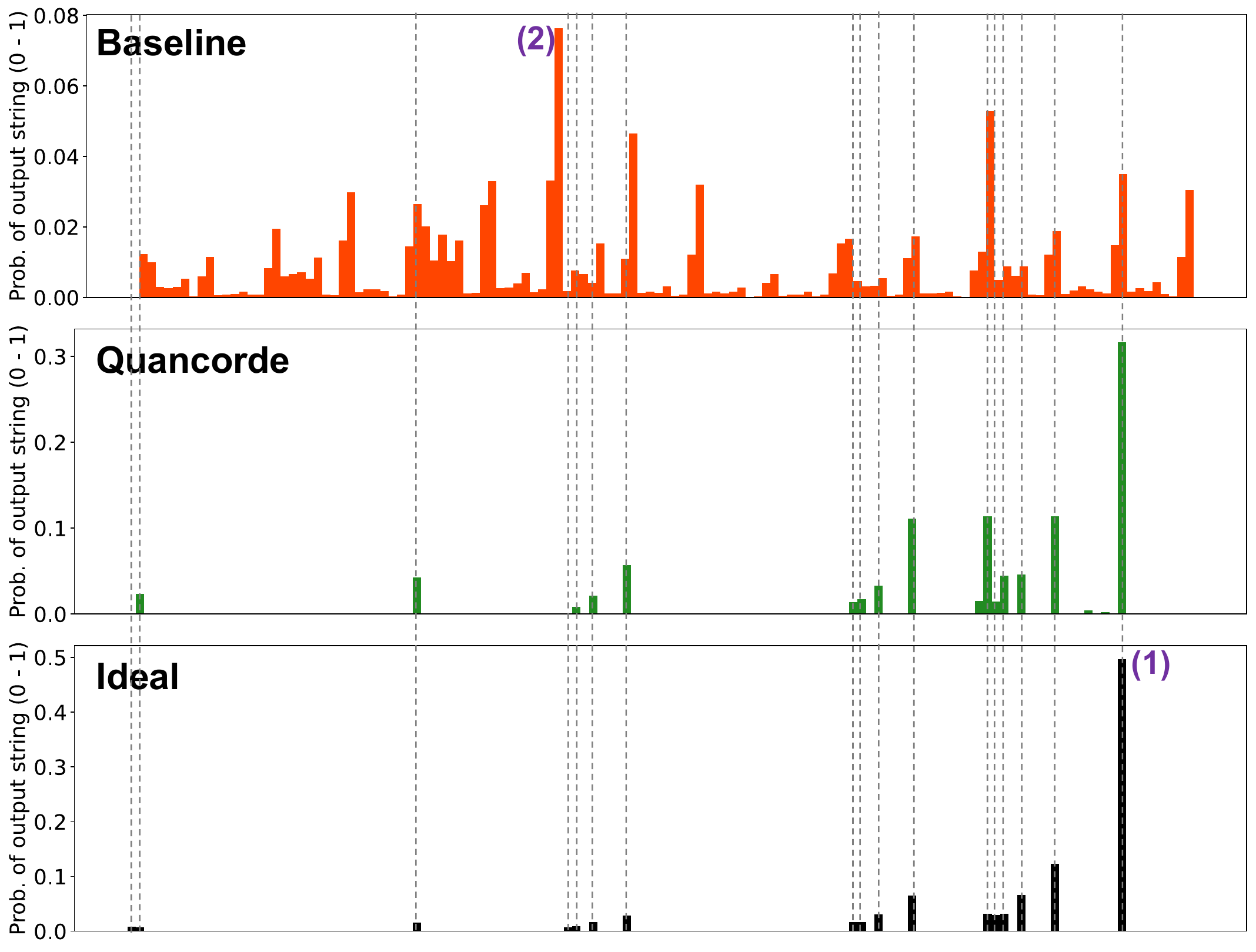}%}
\caption{Quancorde benefits in non-uniformly weighted multi-outcome applications, shown for a 7-qubit circuit with 17 outputs. The Quancorde output distribution closely matches the Ideal distribution, while the baseline distribution does not. Some correct / incorrect peaks are indicated. %with correct high-probability peaks like (1). Whereas, the baseline distribution (orange) has very poor fidelity with incorrect high-probability peaks like (2).
%Quancorde identifies all 17 bitstrings among the top 24 Ranks of the 128 bitstring search space. %TODO CAMERA
} 
\label{fig:quancorde_multi}
\end{figure}

\subsection{Non-uniform Multi-output usecase}
\label{e:multi}

Quancorde can also achieve substantial benefits for applications with non-uniformly weighted multi-outcomes.
Fig.\ref{fig:quancorde_multi} shows Quancorde benefits for a 7-qubit circuit with a correct outcome distribution comprising of a superposition of 17 7-bit strings.
The experiment is run with an intra-machine ensemble on Montreal.
The baseline distribution (orange) is far off from the ideal distribution in purple.
For example, peak (2) in the baseline is absent in the ideal, and other peaks can be similarly inferred. 
Quancorde's correlation processing identified all 17 strings amongst its top 24 Ranks.
The weighted output is shown in green.
Clearly it bears good resemblance to the Ideal with peaks such as (1).
While still imperfect, a fidelity boosting of 4.6x / 2.1x (`Mean' / `vs Best', respectively) is achieved over the baseline.

\iffalse
Quancorde's ability to shortlist the correct outcome(s) into a reasonably sized subset of bitstrings is also useful for  non-uniformly weighted multi-outcome applications with a non-trivial number of correct output bitstrings.

Quancorde is able to identify all 17 strings among its top 24 Ranks, i.e., among the top 24 strings whose output probability ordering across the ensemble are most correlated with the canary-produced ensemble order.
A superposition of these top 24 states (or even conservatively, say the top 30 states) can potentially be used to achieve considerable computational speedup over a naive baseline.
\fi

\begin{figure}[h]
\centering
%\fbox{
\includegraphics[width=0.95\columnwidth,trim={0cm 0cm 0cm 0cm},clip]{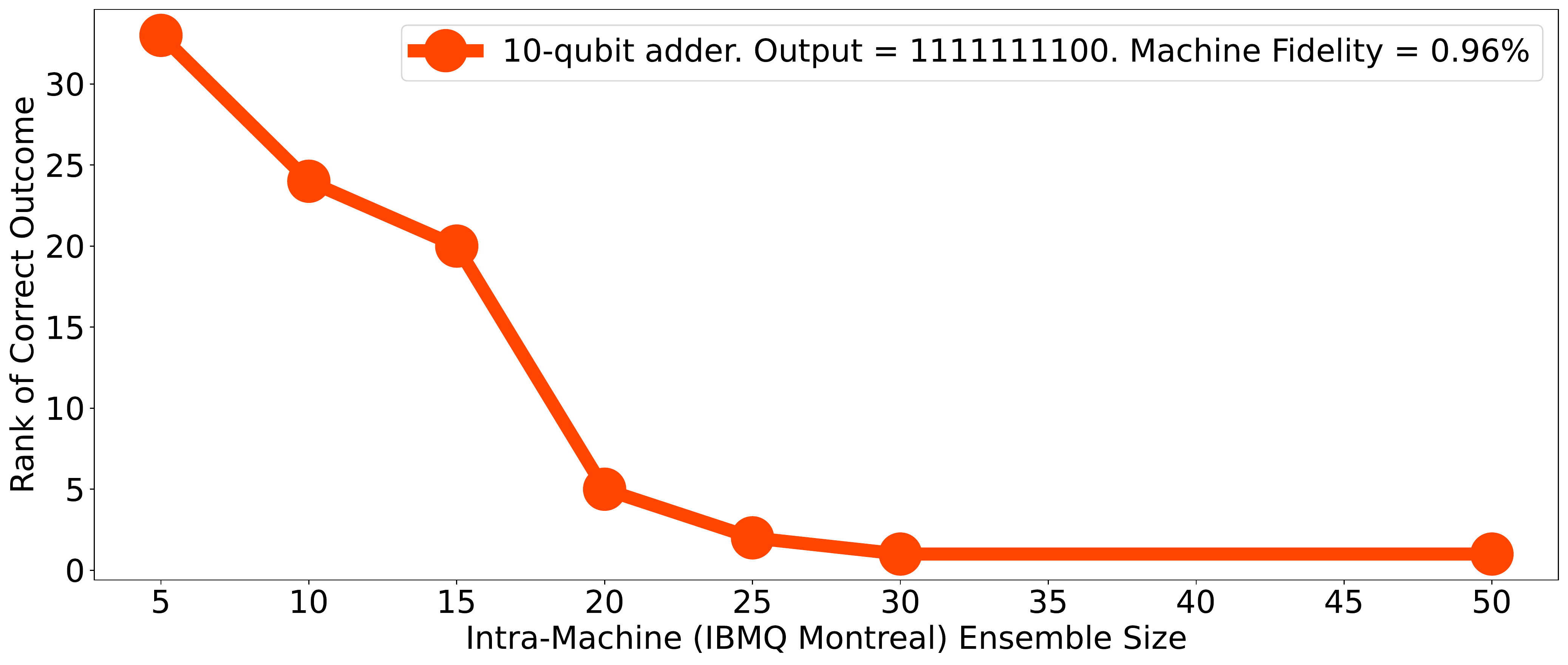}%}
\caption{Rank of correct outcome improves as size of intra-machine ensemble increases, for a 10-qubit adder on IBM Quantum Montreal. 
% Increased diversity (through ensemble size) improves the Quancorde capability to identify the correct outcome. %TODO CAMERA
} 
\label{fig:quancorde_ensemble}
\end{figure}

\subsection{Sensitivity to ensemble size}
\label{e:size}

Greater diversity and larger ensemble sizes could increase the potential for correct outcome identification.
Fig.\ref{fig:quancorde_ensemble} shows how the Rank of the correct outcome for a 10-qubit adder circuit improves as we increase the size of its intra-machine Quancorde ensemble. 
The experiment is performed on IBM Q Montreal.
Even a small ensemble of just five mappings is fairly successful in achieving a top-30 Rank.  
As is evident, increasing the ensemble size improves the Rank.
This is because a large ensemble has more diverse mappings or machines.
This means that there are more diverse noise effects over which the occurrence probabilities of the different output strings are evaluated.
%This allows for a more unique of the diverse noise characteristics on the  becomes more unique.
Consequently, the correlation of the unique ensemble order for the correct outcome, with the canary-produced ensemble order, is relatively magnified, thus boosting fidelity further.
The benefits reach the maximum (Rank = 1) at an ensemble size of 30 mappings.
%Thus, the correct outcome can be identified with greater clarity, via better Ranks (shown) and greater RCs (not shown).
The optimal size of the ensemble is not directly related to the size of the circuit and is more tied to the baseline fidelity.
For instance, as machine quality improves, deeper / wider circuits can sufficiently be boosted by Quancorde, even with small ensembles. 
\section{Discussion}
\label{sec:discuss}

\label{FW}

Section \ref{e:size} showed that increasing ensemble size, effectively increases diversity, and thus improves  Quancorde capability. However, there are resource cost constraints to how large ensembles can get. On the other hand, diversity can also be improved in a more intelligent manner through diversity-focused noise-aware compilation schemes which is a novel direction for future exploration.
%\circled{5}\ While canary circuits are only leveraged for ensemble ordering in this work, they are likely to have other potential uses, especially in quantum resource management, since they are shown to be considerably more indicative of application-machine relations in comparison to state-of-the-art error heuristics.
%Canary circuits are required to be classically simulable - only then are their correct outcomes known and their fidelity on machines can be evaluated. In this work, we employ only purely Clifford canary circuits. 
Prior work has shown that efficient classical simulation can be extended beyond strictly Clifford circuits to constrained Clifford+T circuits wherein T refers to the single-qubit 45-degree phase shift~\cite{Bravyi2016,Bravyi2019}.
Designing canary circuits with a mix of Clifford gates and minimal T gates is worth exploring. %TODO CAMERA

%\cirxled{6} sensiticity to output / encoding schemes to create more uniqueness

%\cirxled{7} undeerstanding why ordering works for very low fidelity

\label{RW}
Quancorde bears some philosophical resemblance to zero-noise extrapolation~\cite{temme2017error,li2017efficient,giurgica2020digital}.
In ZNE, a target quantum circuit is transformed (in a variety of possible ways) to run at different effective levels of quantum device noise. 
Doing so allows the result of the target computation to be computed across the variety of noise levels, which can then be extrapolated to a noiseless level. 
The utility of ZNE is tied to the effectiveness of noise scaling and extrapolation. 
Exploring deeper connections between Quancorde and ZNE is a worthy pursuit.

Clifford circuits have played a critical role in multiple quantum domains such as quantum communications / networks~\cite{Veitch2014}, error correction codes~\cite{QECIntro}, teleportation~\cite{gottesman1999demonstrating}.
%Clifford circuits have also be employed in manners with some similarity to the canary circuits proposed here. 
~\cite{czarnik2020error,strikis2021learningbased} use Clifford circuits  to learn better error mitigation strategies for target circuits. 
Near-Clifford circuits have also been used to improve the efficiency of Dynamic Decoupling~\cite{Das2021}. These techniques can be employed orthogonal to Quancorde.
Further, accreditation protocols~\cite{Ferracin_2021} provide an upper bound on the variation distance between the probability distribution of the experimental outputs of a noisy quantum circuit and its ideal, noiseless counterpart. To do so, random Clifford circuits similar to canary circuits are utilized.
The number of Clifford circuits needed grows considerably with the error tolerance and confidence with which the variation distance is estimated.

Finally, ensembling approaches such as EDM~\cite{Tannu:2019b} and EQC~\cite{eqc} have been explored.
While we compare against EDM, EQC is a scheduling technique for distributed VQA.
Quancorde can be employed by quantum cloud vendors~\cite{IBMQE,AWS,Azure} to effectively manage quantum resources to boost fidelity of applications.
Its potential coordination with existing quantum cloud resource managers~\cite{QCloud,QManager} is worthy of exploration.

\iffalse
EDM~\cite{Tannu:2019b} mitigates vulnerability to correlated errors by using diversity in qubit allocation, thus steering the trials towards making different mistakes. 
While EDM effectively achieves the best mapping a device can offer while avoiding correlated errors, it is still fundamentally restricted by the overall noise characteristics of the device. 
However, Quancorde can achieve fidelity beyond the capabilities of any single device.
EQC~\cite{eqc} creates a quantum ensemble, which dynamically distributes quantum tasks asynchronously across a set of physical devices.
%, and adjusting the ensemble configuration with respect to machine status, thereby avoiding temporal-dependant device-specific noise for variational algorithms. %TODO CAMERA
EQC, being a scheduling technique, is fundamentally different from Quancorde.
\fi

\section{Conclusion}
%Advancing the noisy quantum era to the doorstep of quantum advantage would require us to solve quantum applications of at least 50 qubits with reasonable fidelity.
%However, on today's quantum devices, execution fidelity tends to collapse dramatically for most circuits beyond a handful of qubits.
%Thus, it is also imperative to employ fundamentally novel techniques that are able to boost quantum application fidelity in new ways, beyond the limitations of today's noisy quantum devices. %TODO CAMERA

Quancorde is a novel approach to identifying the correct outcomes of extremely low fidelity quantum applications by exploiting quantum device diversity.
It uses  Clifford canary circuits to order a diverse ensemble of devices or qubits/mappings along the direction of increasing fidelity of the target application.
It then identifies the correlation of the application outputs' ensemble ordering with the canary order, and uses this to weight the application output distribution, resulting in boosted application fidelity.
Quancorde is especially useful when diversity is significant, and applications' ideal outcomes are hard to produce - both of which are intuitive expectations in the coming years of the noisy quantum era.
Quancorde can have a revolutionary impact on the way noisy quantum resources are effectively leveraged to boost the fidelity of real-world quantum use cases.

\section*{Acknowledgement}
This work is funded in part by EPiQC, an NSF Expedition in Computing, under award CCF-1730449; 
in part by STAQ under award NSF Phy-1818914; in part by NSF award 2110860; 
in part by the US Department of Energy Office  of Advanced Scientific Computing Research, Accelerated Research for Quantum Computing Program; 
and in part by the NSF Quantum Leap Challenge Institute for Hybrid Quantum Architectures and Networks (NSF Award 2016136) 
and in part based upon work supported by the U.S. Department of Energy, Office of Science, National Quantum Information Science Research Centers.  
This work was completed in part with resources provided by
the University of Chicago's Research Computing Center.
GSR is supported as a Computing Innovation Fellow at the University of Chicago. This material is based upon work supported by the National Science Foundation under Grant \# 2030859 to the Computing Research Association for the CIFellows Project.
KNS is supported by IBM as a Postdoctoral Scholar at the University of Chicago and the Chicago Quantum Exchange.
FTC is Chief Scientist for Quantum Software at ColdQuanta and an advisor to Quantum Circuits, Inc.

%%%%%%%%% -- BIB STYLE AND FILE -- %%%%%%%%
\bibliographystyle{IEEEtranS}
\bibliography{refs}
%%%%%%%%%%%%%%%%%%%%%%%%%%%%%%%%%%%%

\end{document}